\documentclass[12pt]{article}

\usepackage{slashed}
\usepackage{color, verbatim}
\usepackage{latexsym}
\usepackage{epsfig}
\usepackage{amsmath,accents}

\newcommand*{\dt}[1]{%
  \accentset{\mbox{\scriptsize\bfseries .}}{#1}}

\usepackage{amssymb}
\usepackage{graphicx}
\usepackage{bm}
   
\makeatletter

\@addtoreset{equation}{section}
\makeatother

\newcommand{\be}{\begin{equation}}
\newcommand{\ee}{\end{equation}}
\newcommand{\bea}{\begin{eqnarray}}
\newcommand{\eea}{\end{eqnarray}}

\newcommand{\h}{\mathfrak h}

\newcommand{\vev}[1]{\left\langle #1\right\rangle}

\begin{document}

\thispagestyle{empty}

\begin{center}

\hfill MPP-2015-213  \\
\hfill UAB-FT-768
\begin{center}

\vspace{.5cm}

{\Large\sc On dilatons and the LHC diphoton excess}

\end{center}

\vspace{1.cm}

\textbf{
Eugenio Meg\'ias$^{\,a}$, Oriol Pujol\`as$^{\,b}$, Mariano Quir\'os$^{\,b\,,c\,,d}$
}\\

\vspace{1.cm}
${}^a\!\!$ {\em {Max-Planck-Institut f\"ur Physik (Werner-Heisenberg-Institut), \\ F\"ohringer Ring 6, D-80805, Munich, Germany}}

\vspace{.1cm}
${}^b\!\!$ {\em {Institut de F\'{\i}sica d'Altes Energies (IFAE),\\ The Barcelona Institute of  Science and Technology (BIST),\\ Campus UAB, 08193 Bellaterra (Barcelona) Spain}}

\vspace{.1cm}
${}^c\!\!$ {\em {Instituci\'o Catalana de Recerca i Estudis  
Avan\c{c}ats (ICREA), \\ Campus UAB, 08193 Bellaterra (Barcelona) Spain}}

\vspace{.1cm}
${}^d\!\!$ {\em {ICTP South American Institute for Fundamental Research \& \\ Instituto de F\'isica Te\'orica, Universidade Estadual Paulista, \\ 
S\~ao Paulo, Brazil}}

\end{center}

\vspace{0.8cm}

\centerline{\bf Abstract}
\vspace{2 mm}
\begin{quote}\small
We study soft wall models that can embed the Standard Model and a naturally light dilaton.
Exploiting the full capabilities of these models we identify the parameter space that allows to pass Electroweak Precision Tests with a moderate Kaluza-Klein scale, around $2$ TeV.
We analyze the coupling of the dilaton with Standard Model (SM) fields in the bulk, and discuss two applications: i) Models with a light dilaton as the first particle beyond the SM pass quite easily all observational tests even with a dilaton lighter than the Higgs. However the possibility of a 125 GeV dilaton as a Higgs impostor is essentially disfavored; ii) We show how to extend the soft wall models to realize a 750 GeV dilaton that could explain the recently reported diphoton excess at the LHC. 
\end{quote}

\vfill

\newpage

\tableofcontents

\newpage

 \section{Introduction}
 \label{introduction}

After the recent discovery of a Higgs-like resonance at LHC, it is of the greatest importance to assess what kind of Higgs-like particle it is. 
One of the most interesting alternatives to the Standard Model (SM) Higgs is that this state actually is a dilaton, a pseudo-Goldstone boson of the Spontaneous Breaking of Conformal Invariance (SBCI)~\cite{Rattazzi:2000hs,Goldberger:2008zz}. This pattern of symmetry breaking forces that the dilaton couplings to fermions/gauge bosons are proportional to their mass and therefore a dilaton could in principle well mimic a SM Higgs.
Recent progress indicates that a light dilaton can indeed emerge naturally in strong (or extra dimensional) sectors~\cite{CPR}, see also~\cite{Bellazzini:2013fga,Coradeschi:2013gda,Serra:2013kga,Megias:2014iwa,Cox:2014zea,Megias:2015nya}.  
Models of this type can embed the Standard Model (SM) and might give a concrete realization of a light dilaton. 

On the experimental side, discriminating between a SM Higgs and a Higgs plus a light dilaton is not straightforward because it requires an accurate measurement of the `Higgs' couplings. The current uncertainty in these measurements being around 10\%~\cite{combination}, the pessimist will find it as little indication beyond the SM. Seeing the bottle half-full, though, one might say that there is still a considerable room to fit such an option (at least while awaiting LHC Run II results). 

The available phenomenological studies on light dilatons (and even on dilaton impostors) are mostly based either on Effective Field Theory (EFT) analyses or on rather narrow classes of five-dimensional models~\cite{Goldberger:2008zz,Bellazzini:2013fga,Coradeschi:2013gda,Serra:2013kga,Appelquist:2010gy,Grinstein:2011dq,Vecchi:2010gj,Low:2012rj,Chacko:2012sy,Chacko:2012vm,Bellazzini:2012vz,Chacko:2013dra}. 
The structure of the dilaton interactions essentially follows from the SBCI pattern, but this still leaves a significant amount of model dependence. In the first place, the overall strength of the dilaton couplings depends on a new parameter, $f$, the scale of the SBCI which does not need to exactly coincide with the electroweak (EW) scale $v=246$~GeV. Just like for the SM Higgs, the trilinear dilaton couplings are proportional to the mass: for fermions (gauge bosons) they  scale like $m_F/f$ ($m_V^2/f$). Thus, if $f$ is close to, but not exactly, $v$ then the dilaton can look quite close to the SM Higgs. The main challenge, then, is how to comply with the rest of observations such as the EW precision tests (EWPT). We discuss this below, but it is not hard to imagine that considering a rich enough class of models can help in this respect. 
Let us now add that there is actually more model-dependence coming from the fact that naturally light dilatons require  some amount of explicit breaking of conformal invariance~\cite{CPR}, which can be introduced in a number of ways. The dilaton couplings to the various SM species are sensitive to how much explicitly/spontaneously that species is breaking conformal invariance. 
Then for the present discussion it seems relevant to analyze, within some well defined class of models, 
whether or not there is any viable dilaton model that looks like a SM Higgs within 10\% of the various measured couplings.
 
Having this as our main motivation, we aim to study in this article the simplest workable models that naturally produce light dilatons, in the form of soft walls, or warped extra dimensions. We consider a class of models that is simple enough to be tractable and large enough to include the necessary \textit{dials} that allow to exploit the full capabilities of extra-dimensional models. This is crucial in order to possibly realize realistic values of the dilaton couplings to the rest of the SM particles as well as to pass the experimental tests.

For instance, concerning the dilaton as a Higgs-impostor application, in the extra-dimensional realizations the Kaluza-Klein (KK) scale $m_{KK}$ (the mass of the first KK-resonance) must not exceed a few TeV, simply because $m_{KK}$ and $f$ (that is, the EW scale $v$) are then linked by $m_{KK} \lesssim 4\pi v$. This already poses an important constraint, since in too simplified models even EWPT are enough to push $m_{KK}$ up to around $10$ TeV, which would render the construction unattractive. Fortunately we find that soft wall models naturally contain the ingredients that allow a low $m_{KK}$ and pass EWPT. Technically, this results from having superpotentials with exponential dependence on the Goldberger-Wise scalar $\phi$ (see also {\em e.g.}~Ref.~\cite{Elander:2015asa} for recent analyses of similar models) in the infrared. This is known in the literature as \textit{hyperscaling} and it seems to correspond in the conformal field theory (CFT) interpretation to a quasi-scale invariant regime that is parameterized by the exponent in the superpotential. We will not enter here into the full meaning and CFT interpretation of this hyperscaling like regime. We simply observe that it is one of the \textit{dials} that the models will benefit from.

Another important constraint concerns naturalness. It has been recently appreciated that the SBCI, and therefore the appearance of a light dilaton in the spectrum, can be realized in a natural way~\cite{CPR,Bellazzini:2013fga,Coradeschi:2013gda,Serra:2013kga,Megias:2014iwa,Cox:2014zea,Megias:2015nya}, 
and the simplest realizations can keep the dilaton lighter than the SBCI scale by up to a factor 10-100. Provided  $m_{KK}$  can be kept around the few TeV range, this should allow to comfortably make a light dilaton and even a dilaton Higgs impostor candidate from this respect.  

Besides the possibility of a hypothetical Higgs-impostor scenario, there are at least two more reasons why light dilaton models are interesting/useful (beyond the fact that the dilaton represents the lightest state of a quite large class of strong/extra-dimensional models). 
Adding a light dilaton-like scalar to the SM Higgs has a very positive impact on baryogenesis~\cite{Servant:2014bla}. 
Lastly, the recently reported diphoton excess at LHC~\cite{excess} could be modelled as due to a $\sim$~750~GeV dilaton. Since the dilaton is naturally expected to be the lightest excitation, this possibility has a certain added appeal. In summary, our aim is to assess whether warped extra dimensional models are able to model a naturally light dilaton, or it can exist a viable 750~GeV dilaton giving rise to the $\gamma\gamma$  ATLAS and CMS excess.

Let us now briefly sketch the soft wall models that we shall consider here, and how they look like in the CFT interpretation. We will consider that the five-dimensional (5D) bulk geometries are close to Anti-de-Sitter (AdS) near the `ultraviolet (UV) brane', and that they are driven away from AdS by a neutral scalar field $\phi$ near the `infrared (IR) brane'.  This picture is dual to a (UV fixed point) CFT deformed by a scalar singlet operator $\cal O$ (dual to $\phi$), in such a way that the deformation is confining -- it produces the infrared scale $m_{KK}$. The dilaton that appears in these models arises from the $\cal O$ operator -- it is one of its excitation modes. In addition to this, we include 5D versions of SM fields, including a 5D Higgs doublet. 
In the CFT picture, the CFT contains SM-like operators, including a Higgs doublet $H$. 
The breaking of conformal invariance and of the EW symmetry are naturally related because of the allowed couplings between $H$ and $\phi$.
Upon breaking EW symmetry, there can be up to two types of resonances: one from  $\cal O$ and the other from the radial direction of $H$. 
 Our main aim here is to show how to construct  models of this sort which are phenomenologically viable and  based on the well understood tools from extra dimensional models.

This paper is organized as follows. In Sec.~\ref{model} we introduce the extra dimensional model, and present the results for the dilaton mass and the KK spectrum of vector and tensor fluctuations. A semi-analytical approximation for these results can be obtained from some mass formulae, which allows the computation of the lightest modes of the spectra by simple integrals of the background fields. Some versions of these formulae will be also presented in this section. We confront these results with EWPT, and find the corresponding bounds for the dilaton and the KK masses in Sec.~\ref{sec:EWPT}. We study in Sec.~\ref{sec:coupling} the coupling of the light dilaton with SM matter fields which essentially disfavors the possibility of a 125 GeV dilaton as a Higgs impostor. We analyze in Sec.~\ref{sec:diphoton} an extension of this model and its implications on the couplings, and the application to the diphoton excess by a 750 GeV dilaton. Finally we conclude with a discussion of our results in Sec.~\ref{sec:conclusions}. Some technical details are left to Appendix~\ref{sec:AppendixA}.

\section{Light dilatons from an extra dimension} 
\label{model}

In this section we describe generic 5D warped models that give rise to a naturally light dilaton state. In the subsections we particularize these results to a specific benchmark model, and study in details its background properties and the corresponding spectrum of excitations.
We consider a 5D space with an arbitrary metric $A(y)$ such that in proper coordinates
\be
ds^2=e^{-2 A(y)}\eta_{\mu\nu}dx^\mu dx^\nu+dy^2\,,
\label{ds2}
\ee
where $\eta_{\mu\nu}=(-1,1,1,1)$, and two branes localized at $y=y_0=0$ and $y=y_1$ which we will refer in the following as UV and IR branes respectively. We are following the conventions of Refs.~\cite{Cabrer:2009we,Cabrer:2010fb,Cabrer:2011fb,Brouzakis:2013gda}.

The dynamics of the coupled scalar-gravitational system is defined by the action
\begin{align}
S&=M^3\int d^5x\sqrt{-g}\left(R-\frac{1}{2}(\partial_M \phi)^2
-V(\phi)\right) \nonumber \\
&-M^3\sum_{\alpha}\int d^4x dy \sqrt{-g}\,2\mathcal V^\alpha(\phi)\delta(y-y_\alpha)\,,
\label{action}
\end{align}
where $\mathcal V^\alpha$ ($\alpha=0,1$) are the four-dimensional (4D) brane potentials and $M$ is the 5D Planck scale. The dilaton field $\phi$ in the above action is dimensionless, while the mass dimension is $2$ for $V(\phi)$ and $1$ for $\mathcal V_\alpha(\phi)$. The 4D (reduced) Planck mass $M_{Pl}=2.4\times 10^{18}$ GeV is related to $M$ by~\footnote{In what follows we will set units where $M^3=1$, and define $\dot{X}(y)\equiv dX/dy$, $X'(\phi)\equiv dX/d\phi$.}
\be
M_{Pl}^2=2 M^3 \int e^{-2A}dy\,.
\label{relacion}
\ee
The equations of motion are
\begin{align}
3\ddot{A}(y)&=\frac{\dot{\phi}^2(y)}{2}+\sum_\alpha \mathcal V_\alpha(\phi) \delta(y-y_\alpha) \,, \\
6\dot{A}^2(y)&=-\frac{V(\phi)}{2}+\frac{1}{4}\dot{\phi}^2(y) \,, \\
\ddot{\phi}(y)-4\dot{A}(y)\dot{\phi}(y)&=V'(\phi)+\sum_\alpha2\mathcal V'_\alpha(\phi) \delta(y-y_\alpha) \,.
\end{align}
The boundary conditions on the branes are obtained by integrating in a small interval around $y=y_\alpha$, yielding
\be
\left.\dot{A}(y)\right|^{y_\alpha^+}_{y_\alpha^-}=\frac{1}{3}\mathcal V_\alpha(y_\alpha)\,,\quad
\left.\dot{\phi}(y)\right|^{y_\alpha^+}_{y_\alpha^-}=2\mathcal V'_\alpha(y_\alpha) \,.
\ee
By imposing  $\mathbb Z_2$ orbifold symmetry across each brane one obtains 
\be
\dot{A}(y_\alpha)=(-1)^\alpha\frac{1}{6}\mathcal V_\alpha(y_\alpha)\,, \quad
\dot{\phi}(y_\alpha)=(-1)^\alpha\mathcal V'_\alpha(y_\alpha) \,.
\ee

As is well known, the equations of motion can be reduced to first-order form by introducing a superpotential $W(\phi)$~\cite{DeWolfe:1999cp,Gubser:2000nd}, given by~\footnote{This equation can be viewed as an equation for $V$ given $W$, or the converse. In this section we will consider $W$ as an input, and will focus on the specific model in Section~\ref{sec:bench}.}
\be
 V(\phi) \equiv \frac{1}{2} \left[ W'(\phi) \right]^2 - \frac{1}{3} W(\phi)^2 \,.
\label{VW}
\ee
The background equations of motion then reduce to 
\be
\dot{A}(y) = \frac{1}{6} W(\phi(y)) ,\quad
\dot{\phi}(y) =  W'(\phi)~.
 \label{metricEOM}
\ee
It is then convenient to introduce the localized effective potentials
\be\label{U}
U_\alpha(\phi)\equiv\mathcal V_\alpha(\phi)-(-1)^\alpha W(\phi) \,.
\ee
With these, the boundary conditions together with the equations of motion lead to 
\be
U_\alpha(\phi)\big|_{y=y_\alpha} =0 ,\quad
U_\alpha^\prime(\phi)\big|_{y=y_\alpha}=0\ .
\label{BCs}
\ee
Thus one can think that the brane dynamics $\mathcal V^\alpha(\phi)$ fixes the values of the field on the brane, which we shall call \mbox{$\phi_\alpha=(\phi_0,\,\phi_1)$} for the UV and IR branes respectively~\footnote{One can trade the values of $\mathcal V'_\alpha(\phi_\alpha)$ by the branes' locations $y_\alpha$ (or $\phi_\alpha$). Equation~(\ref{BCs}) appears to imply that we are doing two cosmological constant fine-tunings. For general enough solutions, this is really just one fine-tuning because $W$ satisfies a differential equation and one can play with the associated integration constant. It is only in simplified models with a fixed analytical $W$ that this appears like two tunings that allow to eliminate $\mathcal V^\alpha(\phi_\alpha)$.}. 
Setting $A(y_0)=0$,  the inter-brane distance $y_1$, as well as the location of the singularity at $y_s\equiv y_1+\Delta$ and the warp factor $A(y_1)$, are all functions of the field-values $\phi_\alpha$ on the branes.

An important quantity that controls the properties of the dilaton states is the holographic $\beta$-function, defined by
 \be
 -\beta(\phi)=6\frac{W'(\phi)}{W(\phi)}~.
 \label{beta}
 \ee
For exponential superpotentials behaving in the IR as $W\sim e^{{\bar c}\phi}$, the holographic $\beta$-function goes to a constant, $\beta \sim 6{\bar c}$. Models of this type, with ${\bar c} > \sqrt{3/2}$, are able to give `confinement' (a discrete spectrum in all sectors) even without the presence of the IR brane, see {\em e.g.} \cite{Cabrer:2009we}. With ${\bar c} < \sqrt{3/2}$, they give rise to a continuous spectrum with quasi-scaling (known as `hyperscaling' or `hyperscaling violation') and the IR brane is required in order to have a discrete spectrum. 
Even though the confining soft-wall models (with ${\bar c} > \sqrt{3/2}$) appear simpler, realizing a light dilaton requires $\beta(\phi)$ to jump fast enough from small to $\mathcal O(1)$ values \cite{Megias:2014iwa}~\footnote{This will require to introduce one more parameter as we will see next.}.
Instead, if we accept the presence of the IR brane, then the interesting regime is that where the parameter ${\bar c}$ (controlling the  size of the beta function and the amount of hyperscaling in the IR) is small. In both cases, we are introducing basically a one-parameter ($\bar c$) family of models that generalize in one way or another the previous analyses~\cite{Bellazzini:2013fga,Coradeschi:2013gda}. 

\subsection{Benchmark model}
\label{sec:bench}

Depending on whether one defines the model by specifying the superpotential $W(\phi)$ or the potential $V(\phi)$, Eq.~\eqref{VW} can be viewed either as an algebraic equation for $V(\phi)$ or a first order differential equation for $W(\phi)$. The two prescriptions are qualitatively different because the second option involves an additional integration constant that has to be fixed by some appropriate condition. That option is reviewed for instance in \cite{Megias:2014iwa}. The additional integration constant is identified as the condensate $\vev{\cal O}$ of the operator along the deformation direction ($\delta{\cal L} = \lambda {\cal O}$), and the boundary condition that fixes it is that the IR end of the flow is the least singular possible. 

The other prescription --fixing directly the superpotential-- represents a fine-tuned case in which that condensate is artificially set to vanish. Importantly, even in that case, a dilaton mode can still be present. The physical reason seems to be that, even with $\vev{\cal O}=0$, conformal invariance is broken in the infrared by the dynamics. 
The dilaton continues to correspond to the fluctuation of the condensate,
and whether or not it is light is still controlled by the value of $\beta$ at the threshold where $\beta$ starts to grow. Therefore, it is possible (and convenient) to simplify the model by setting to this class of models where $\vev{\cal O}=0$. This is especially convenient once we allow for the presence of an IR brane, which is dual to the condensate of another CFT operator $\vev{\cal O'}$. 

Thus in the present work we will consider an analytic superpotential model defined by two (positive) real parameters $a$ and $c$~\footnote{Notice that the model with superpotential $W=6k\left(1+e^{\nu\phi/\sqrt{6}}\right)$ introduced in Ref.~\cite{Cabrer:2011fb} is a particular case of the superpotential (\ref{nuestro}) for $a=\nu/\sqrt{6}$ and $c=1$.}  
\be
 W(\phi)=6k\left(1+e^{a\phi}  \right)^{c}  \,, \label{nuestro}
 \ee
 from where the holographic $\beta$-function is given by
 \be
 -\beta(\phi)=6\frac{W'(\phi)}{W(\phi)}=6ac\left[1+e^{-a\phi}  \right]^{-1} \,.
 \label{beta}
 \ee
The parameter ${\bar c} \equiv ca $ determines the value of $-\beta(\phi)$ in the $\phi\to\infty$ limit ($6\bar{c}$) and the parameter $a$ governs the slope of $-\beta(\phi)$ at $\phi=0$, which is given by $3 a \bar{c}$. We prefer to use in the following the parameter $c$ for phenomenological convenience.

\begin{figure}[htb]
\vskip .5cm 
 \begin{tabular}{cc}
 \includegraphics[width=7.8cm]{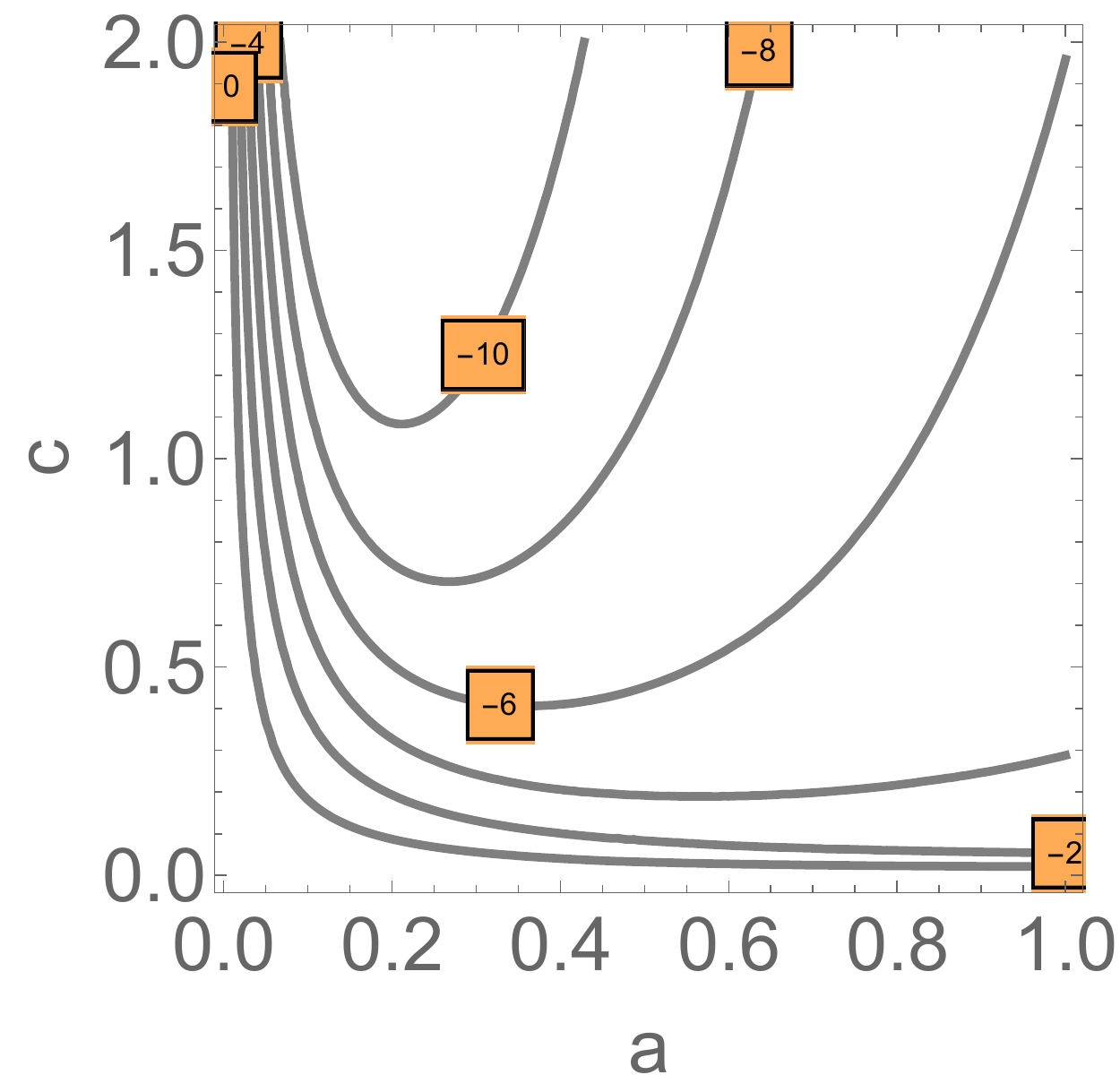}  &
 \includegraphics[width=7.8cm]{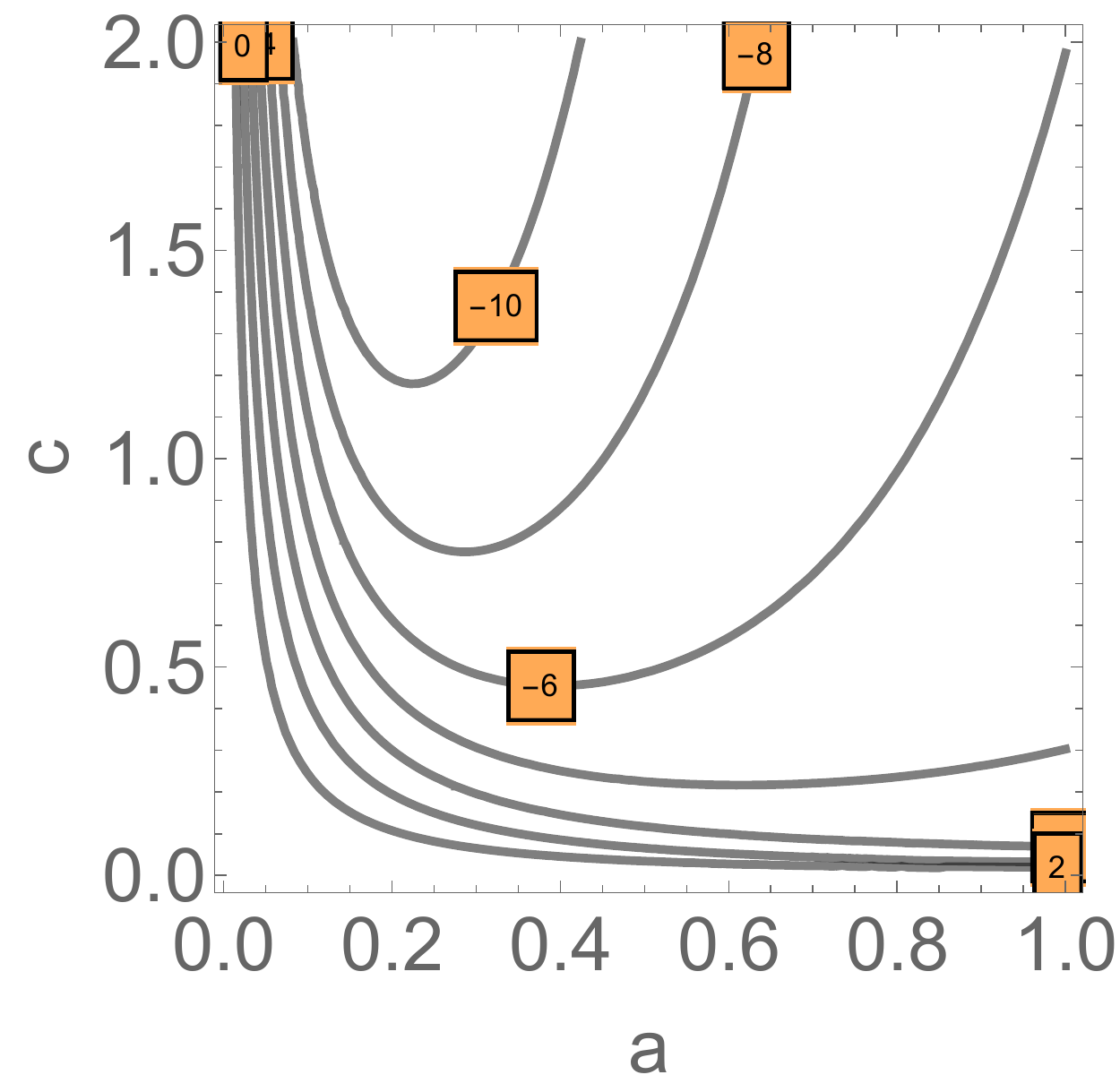}    
\\   
\end{tabular}
\caption{\it Left panel: Contour lines of $\phi_0(a,c,\phi_1=2)$ from the condition $A(\phi_1)=35$. Right panel: Contour lines of $\phi_0(a,c,\phi_1=5)$ from the condition $A(\phi_1)=35$.}
\label{fig1}
\end{figure}

As mentioned above (once the cosmological constant fine-tuning is taken into account) the values of the brane potentials ${\cal V}^\alpha(\phi)$ and their derivatives can be traded by the values of the field \mbox{$\phi=(\phi_0,\,\phi_1)$} on the UV and IR branes respectively. 
One can also write the warp factor, with the condition $A(\phi_0)=0$, as
\be
A(\phi)=B(\phi)-B(\phi_0),\quad B(\phi)=\frac{1}{6ac}\left(\phi-\frac{e^{-a\phi}}{a}\right)  \,.
\label{warp}
\ee
We assume that the brane (potential) dynamics have fixed $(\phi_0,\phi_1)$, such as to solve the hierarchy problem, i.e.~$A(\phi_1)\simeq 35$. This can be done for positive values of $\phi_1$, for which $B(\phi_1)\sim \phi_1$ and negative values of $\phi_0$ such that $B(\phi_0)\sim -e^{-a\phi_0}$. In this way the value of $\phi_0$ exhibits little sensitivity with respects to the value of $\phi_1$ provided the latter is largish. This insensitivity is shown in Fig.~\ref{fig1} where we plot contour lines of constant $\phi_0=(0,-2,-4,-8,-10)$ and $\phi_1=2$ ($\phi_1=5$) in the left (right) panel, after imposing the condition $A(\phi_1)=35$. As we can see, the contour lines have little dependence upon the value of $\phi_1$.
\begin{figure}[htb]
\vskip .5cm 
 \begin{tabular}{cc}  
 \includegraphics[width=7.5cm,height=6.cm]{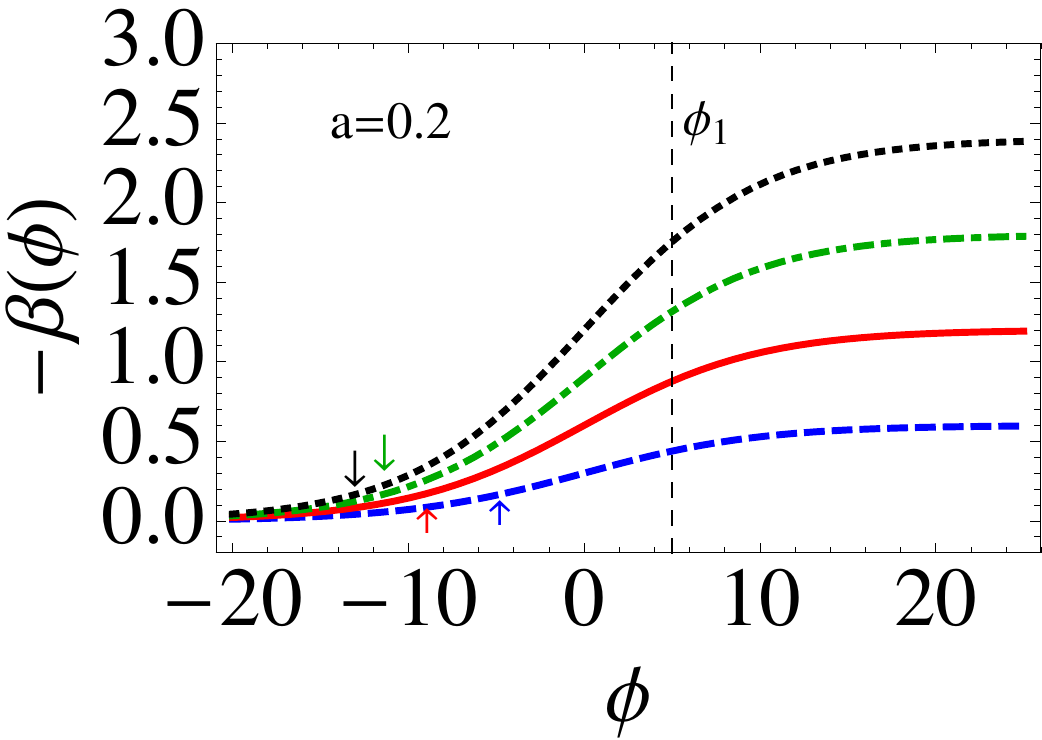}  &
 \includegraphics[width=7.5cm,height=6.cm]{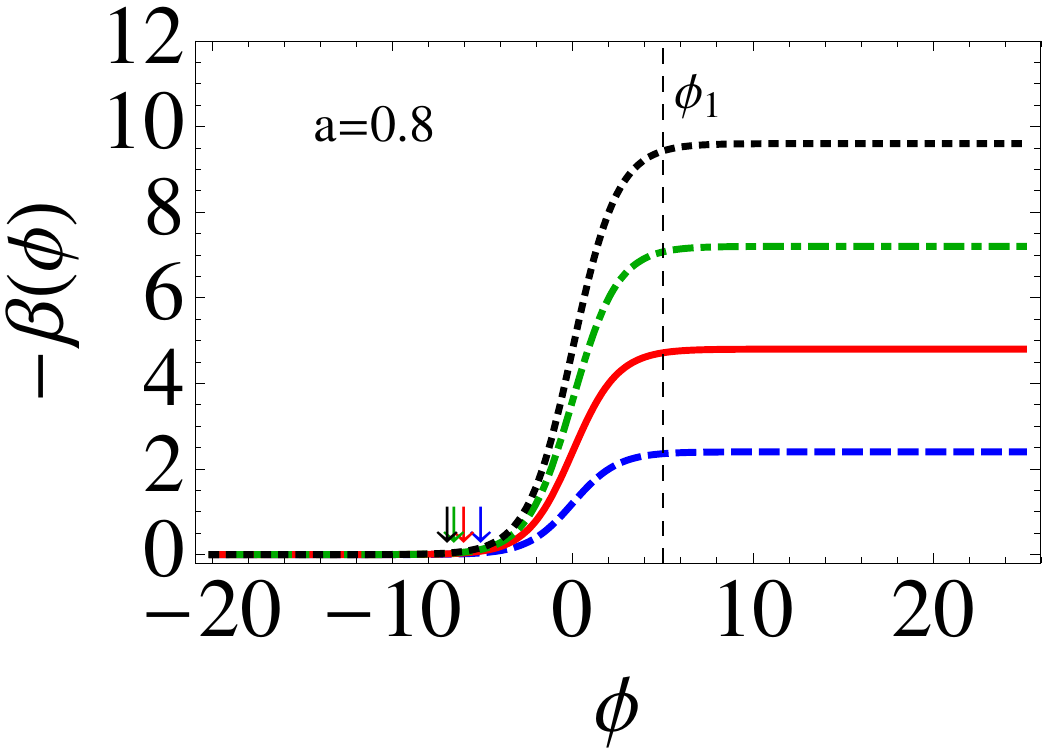}   \\    
\end{tabular}
\caption{
\it The function $-\beta(\phi)$ for $c=$0.5 (dashed blue line), 1 (solid red line), 1.5 (dot-dashed green line) and 2 (dotted black line). We show in the left panel the result with $a=0.2$, and in the right panel with $a=0.8$. The vertical line indicates the position of the IR brane that we are considering throughout the paper, $\phi_1=5$, and the arrows stand for the positions of the UV brane, $\phi_0$, that allow to solve the hierarchy problem, $A(\phi_1) \simeq 35$. $\phi_0$~depends on $c$: from right to left $c=$0.5, 1, 1.5 and 2.
}
\label{fig2}
\end{figure}
We plot then in Fig.~\ref{fig2} the function $\beta(\phi)$ for four different values of $c$=(0.5, 1, 1.5, 2). When using $\phi_1=5$, for $a=0.2$ (left panel) the values of $\phi_0$ fixed by the condition $A(\phi_1)=35$ are $\phi_0=$(-4.8,-8.9,-11.4,-13.0), respectively, and for $a=0.8$ (right panel) one has $\phi_0=$(-5.1,-6.0,-6.6,-6.9), respectively. These are the typical values of the parameters of the model that we are using throughout this work.

The inter-brane distance $y_1$, as well as the location of the singularity at $y_s=y_1+\Delta$ and the warp factor $A(y_1)$, are related to the values of the field $\phi_\alpha$ at the branes. In fact we can use as coordinate the value of the field $\phi$ instead of the value of $y$. The change of coordinates is given by
\be
\frac{dy(\phi)}{d\phi}=-\frac{6}{\beta(\phi)W(\phi)}  \,,
\ee
or, imposing the condition on the IR singularity $y(\phi_s)\equiv y_s$, where $\phi_s\to\infty$
\begin{align}
y(\phi)&=y_s-z(\phi) \,, \nonumber\\
z(\phi)&\equiv\frac{6}{-c\beta(\phi)W(\phi)}\left\{
\frac{c}{a}+\left(\frac{c}{a}-1\right) \textrm{$_2$}F_{1}[1,1,1+\frac{c}{a},-e^{-a\phi}]\right\} \,, \nonumber\\
y_s&=z(\phi_0) \,,
\label{ydephi}
\end{align}
where  $\textrm{$_2$}F_{1}[a,b,c,z]$ is the hypergeometric function defined by
\be
\textrm{$_2$}F_{1}[a,b,c,z]=\sum_{k=0}^{\infty}\frac{a_k b_k}{c_k}\frac{z^k}{k!},\quad
m_k\equiv \Pi_{\ell=0}^k(m+\ell) \,,
\ee
and the last line in Eq.~(\ref{ydephi}) comes from the requirement $y_0=y(\phi_0)=0$. The IR brane is located at $y_1=y(\phi_1)$ which is at a distance $\Delta$ from the singularity $y_s$. They are fixed as:
\begin{align}
y_1&\equiv y_1(a,c,\phi_1)=z(\phi_0(a,c,\phi_1))-z(\phi_1) \,, \nonumber\\
\Delta&= z(\phi_1) \,.
\label{otra}
\end{align}
\begin{figure}[htb]
\vskip .5cm 
 \begin{tabular}{cc}
 \includegraphics[width=7.8cm]{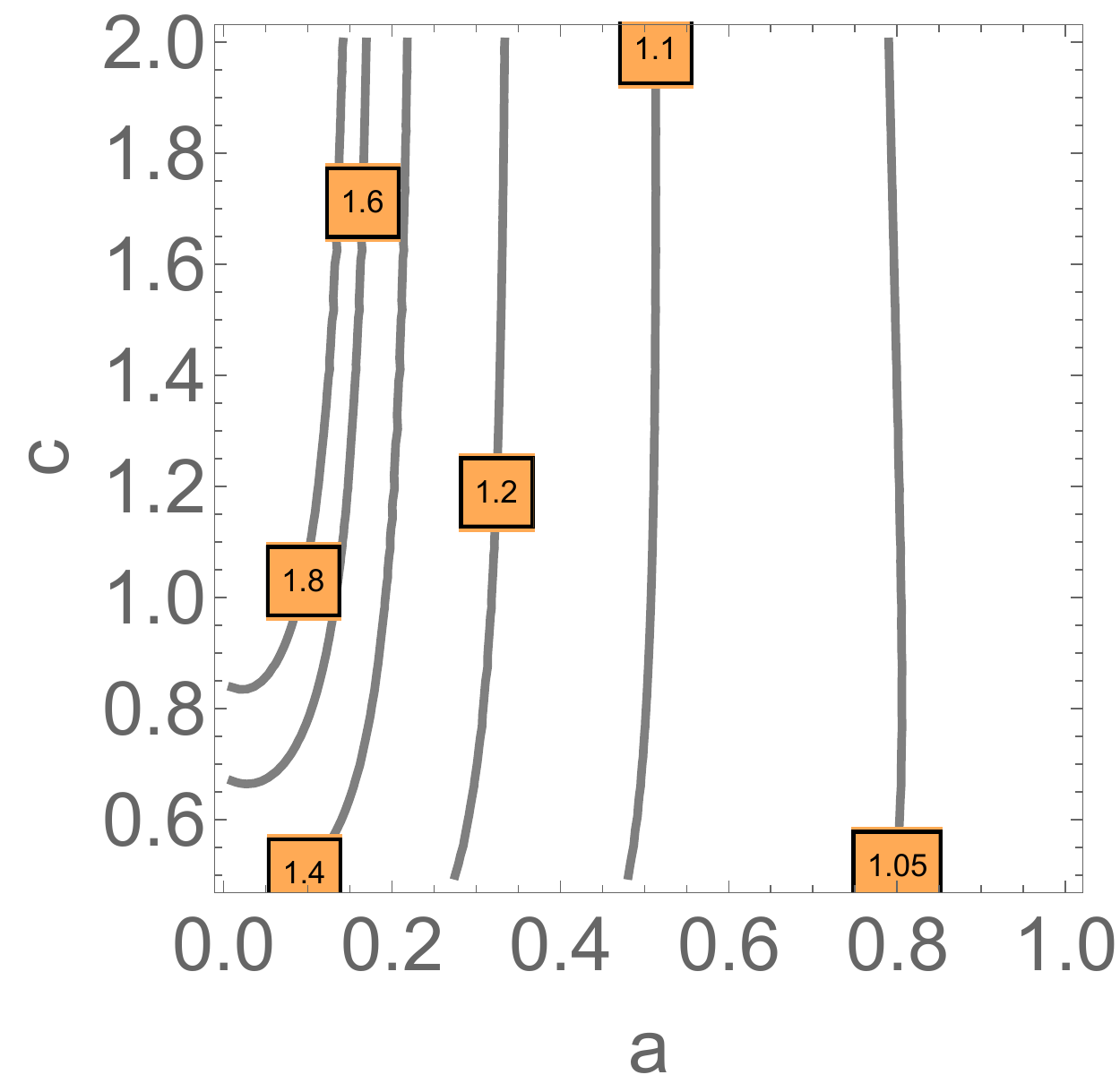}  &
 \includegraphics[width=7.8cm]{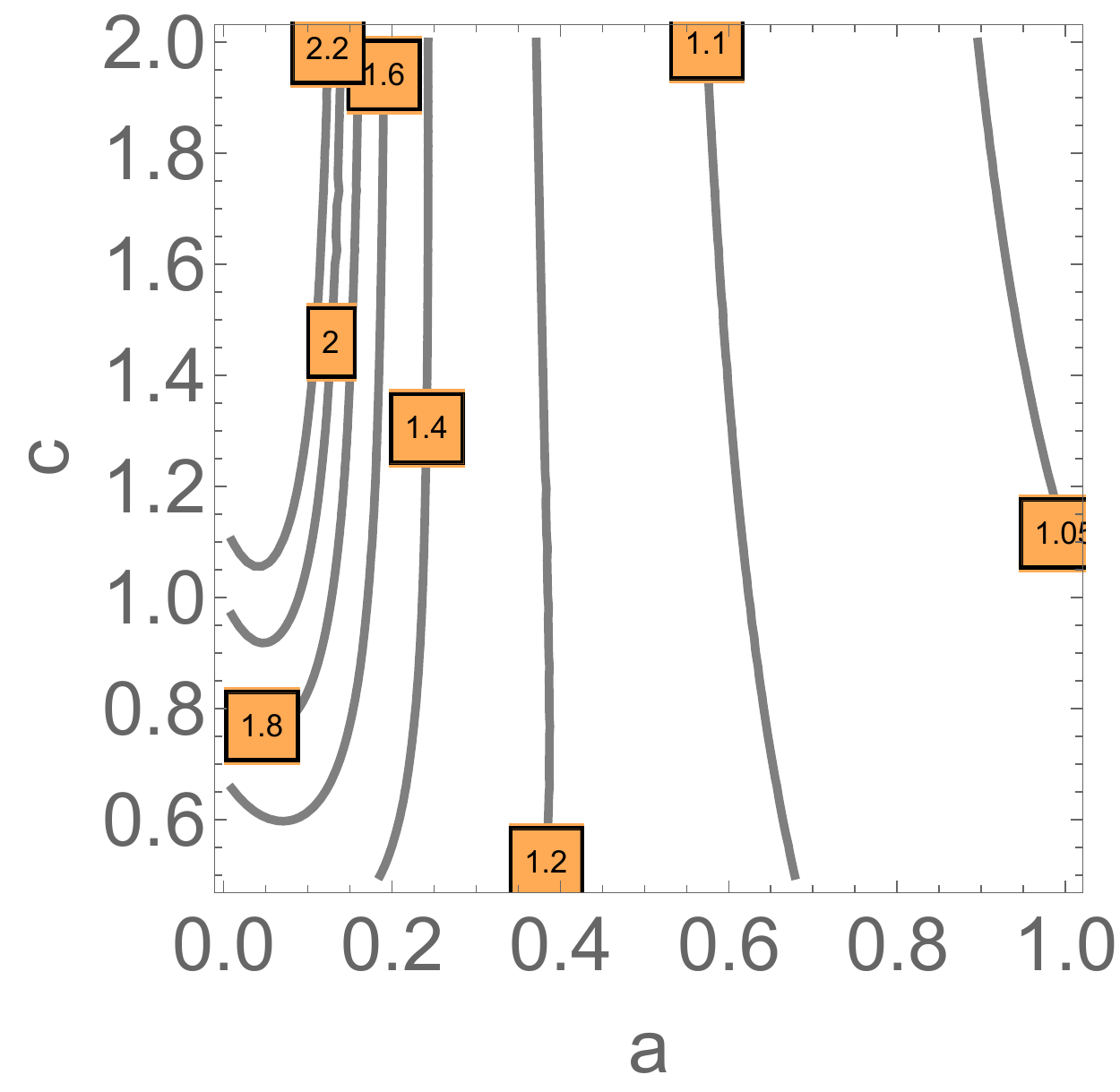}    
\\   
\end{tabular}
\caption{\it Left panel: Contour lines of $A(\phi_1)/ky_1$ from the condition $A(\phi_1)=35$ for $\phi_1=2$. Right panel: Contour lines of  $A(\phi_1)/ky_1$ from the condition $A(\phi_1)=35$ for $\phi_1=5$.}
\label{fig3}
\end{figure}
Contour lines of $A(\phi_1)/ky_1(a,c,\phi_1)$, are shown in the left (right) panel of Fig.~\ref{fig3} for $\phi_1=2$ ($\phi_1=5$). As we can see the larger the value of $\phi_1$ the smaller the value of $y_1$ (and the larger the reduction of the volume). Because of that and the fact that $\phi_0$ has little dependence on $\phi_1$, we will choose from here on the value $\phi_1=5$ and keep $(a,c)$ as free parameters.

Once we have studied the properties of the background in the benchmark model, our next task is to compute the fluctuations around this background and the corresponding spectrum. This will be done in the next subsection.

\subsection{Spectra and mass formulae}

We will present in  this section the basic formulae to compute the spectrum of the scalar, vector and tensor perturbations in the benchmark model. It is not intended to provide a detailed derivation and, when appropriate, we will refer the reader to existing references.

\subsubsection{Scalars}
\label{subsec:massformula}

In order to compute the scalar spectrum and in particular its lightest mode, which is known as the radion/dilaton, we have to consider a scalar perturbation of the background flow solution as~\cite{Csaki:2000zn,Csaki:2007ns}
\begin{align}
ds^2&=e^{-2A(y)-2F(x,y)}\,\eta_{\mu\nu}dx^\mu dx^\nu +[1+G(x,y)]^2dy^2\nonumber\\
\Phi(x,y)&=\phi(y)+\varphi(x,y)
\label{dilatonexcitations}
\end{align}
where the three scalars $F$, $G$ and $\varphi$ are not independent functions but satisfy the relations
\begin{align}
\varphi(x,y)&=\varphi(y)\cdot \mathcal R(x),\quad
F(x,y)=F(y) \cdot \mathcal R(x)\nonumber\\
\varphi(y)&=6\frac{\dot F(y)-2 \dot A(y) F(y)}{\dot\phi(y)},\quad G(x,y)=2 F(x,y)
\end{align}
where we indicate the mode expansion as: $A(y)\cdot B(x)\equiv \sum_n A^{(n)}(y)B^{(n)}(x)$. 

Using the background EOM one can recast the bulk equation of motion and boundary condition for the excitation $F$ as
\begin{align}
& \left(e^{2A}(\ddot A)^{-1}[e^{-2A}F]^{\ \dt{}}\right )^{\dt{}}+\left(m^2 e^{2A}(\ddot A)^{-1}-2 \right)F=0
\label{eq:xi1}\\
&\left.\left[m^2 F+U_\alpha^{\prime\prime}[e^{-2A}F]^{\, \dt{}}   \right]\right|_{y_\alpha}=0
\label{eq:bcUVIRunitary}
\end{align}
\begin{figure}[htb]
 \begin{tabular}{cc}
\includegraphics[width=7.5cm]{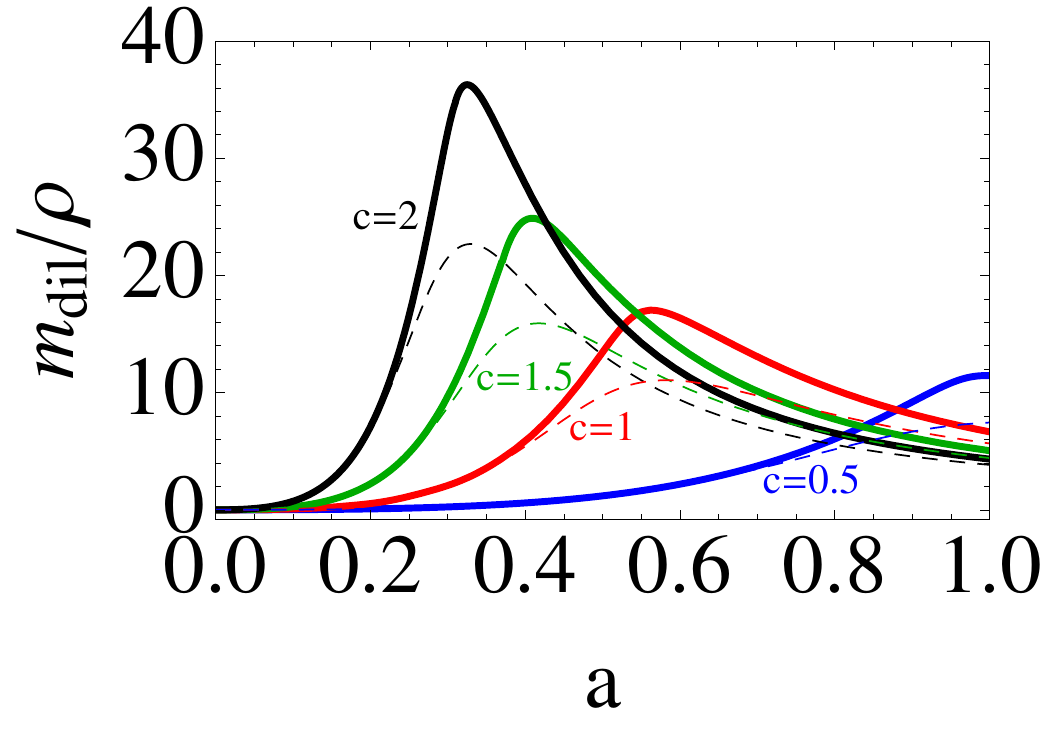}   & 
 \includegraphics[width=7.5cm]{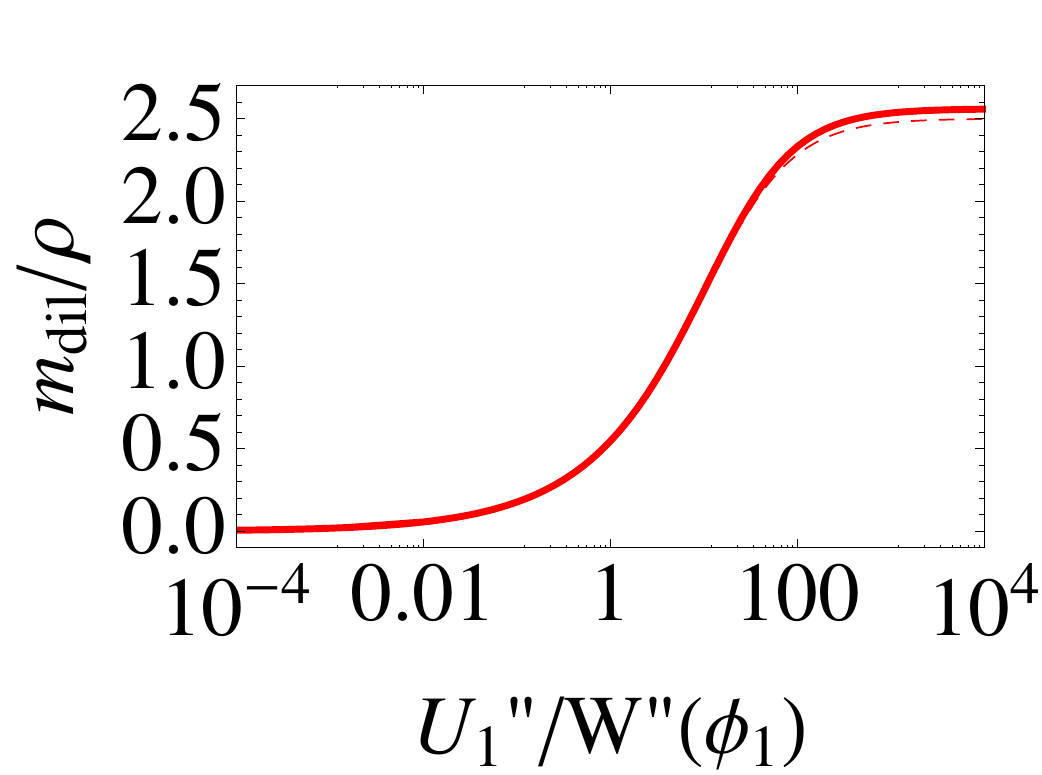} \\  
   \end{tabular}
\caption{\it
Left panel: Mass of the dilaton as a function of the parameter~$a$, normalized to $\rho$. We show the results for $c=0.5, 1, 1.5$ and $2$. We have used $\phi_1=5$, and have considered the values of the potential in the branes $U^{\prime\prime}_0 = W^{\prime\prime}(\phi_0)$ and $U^{\prime\prime}_1 = W^{\prime\prime}(\phi_1)$. Right panel: Dilaton mass as a function of the IR brane potential. The natural value for $U^{\prime\prime}_1$ is $U^{\prime\prime}_1 \sim W^{\prime\prime}(\phi_1)$ as explained in Sec.~\ref{subsec:massformula}. In this plot we have used $c=1$, $a=0.2$ and $U^{\prime\prime}_0 = W^{\prime\prime}(\phi_0)$, and have considered $\phi_1=5$. In both plots we display as wide solid lines the result from a numerical solution of the equations of motion for scalar perturbations,  Eq.~(\ref{eq:xi1}), with the boundary conditions Eq.~(\ref{eq:bcUVIRunitary}), and as narrow dashed lines the result from the mass formula, Eq.~(\ref{eq:mEOuvir}).}
\label{fig4}
\end{figure}
We show in Fig.~\ref{fig4} (left panel) the dilaton mass $m_{\rm dil}$ obtained from a numerical solution of Eq.~(\ref{eq:xi1}) with the boundary conditions Eq.~(\ref{eq:bcUVIRunitary}). These results are not yet in physical units, and we find it convenient to normalize them by the factor 
$\rho \equiv ke^{-A(y_1)}$ which is related to the KK scale $m_{KK}$, see e.g.~\cite{Cabrer:2009we}. An important property is that the value of all these curves is small for small values of $a$, and they tend to zero eventually in the limit $a \to 0$, so that we recover the expected result from an unperturbed AdS metric. These curves exhibit a maximum at around $a \simeq 0.6/c$, which implies the existence of a change of regime: the realization of the dilaton for $a \gtrsim 0.6/c$ is dominated by the $\beta$ function, while for $a \lesssim 0.6/c$ it is dominated by the IR brane. We will call them {\it soft} and {\it hard dilaton} respectively. The soft realization of the dilaton happens when the $\beta$-function is small at the condensation scale and reach confinement fast enough. This is precisely the behavior of the $\beta$-function for large values of $a$, see Fig.~\ref{fig2}. This will be explained in more detail in Sec.~\ref{subsec:massformula}. 

We show in Fig.~\ref{fig4} (right panel) the dependence of the dilaton mass with the value of the IR brane potential. As it will be discussed in Sec.~\ref{subsec:massformula}, the (non fine-tuned) natural value corresponds to $U^{\prime\prime}_1(\phi_1) \sim W^{\prime\prime}(\phi_1)$. Note that this leads to a smaller value for the dilaton mass as compared to the simple choice $U^{\prime\prime}(\phi_1) \to +\infty$~\footnote{This is the choice adopted in e.g. Ref.~\cite{Cabrer:2011fb}. We have checked that we reproduce the results presented in that reference when considering the limit $U^{\prime\prime}(\phi_1) \to +\infty$, and for $a = \nu/\sqrt{6}$ and $c=1$.}.


\noindent\underline{\textit{Mass formulae and types of dilatons}}

It is possible to obtain a general mass formula for the lightest scalar mode -- the dilaton. Details on how to obtain it are deferred to a separate article~\cite{future}. In presence of both UV and IR branes, and with full inclusion of the backreaction, the final mass formula reads
\begin{equation}
\frac{\rho^2}{m_{\rm dil}^{2}} \simeq   
\Pi_2
+  \, \frac{k\, \Pi_0}{U_0^{\prime\prime}\beta_0^2}  
+ \frac{36\,k^2}{U_1^{\prime\prime} \beta_1^2 W_1} 
  \, , \label{eq:mEOuvir}
\end{equation} 
with
\begin{eqnarray}
\Pi_0 &=& ke^{-2A_1}\int_{0}^{y_1} dy \; e^{-2A}\beta^2 \,, \cr
\Pi_2 &=&  k^2 \int_{0}^{y_1} dy \;\frac{e^{4(A-A_1)}}{\beta^2} \left(\frac{36}{W_1}+ \int_y^{y_1} d{\bar y} e^{-2(A-A_1)}\beta^2 \right)\,.
\end{eqnarray}
This formula unifies those found in Refs.~\cite{Cabrer:2011fb} and \cite{Megias:2014iwa} (it slightly differs from that presented in~\cite{Cox:2014zea} in presence of the UV brane). It also recovers the expressions found in~\cite{Csaki:2000zn,Rattazzi:2000hs,Bellazzini:2013fga,Coradeschi:2013gda} when the dominant term in the right-hand side comes from the IR brane. 

The structure of this formula clearly reveals under what conditions a light dilaton emerges. Technically, this happens whenever the right-hand side is large, and one can distinguish three cases when this can be so: from the UV brane, from the IR brane or from somewhere in between. 
In the extreme cases where $U''=0$, it produces an exactly massless mode. This corresponds to the brane world models where the brane potentials are tuned to the bulk superpotential, for which it is well known that there are massless moduli corresponding to the positions of each brane (see e.g.~\cite{Garriga:2001ar,Flachi:2003bb}). Without supersymmetry though, these are clearly tuned models.

To have an idea of how large/small each term can be, note that the structure of the effective potential on each brane is a sum of two terms ($U_{UV}\equiv {\cal V}_{UV} - W$ and $U_{IR}\equiv {\cal V}_{IR} + W$). Therefore a clear notion of tuning emerges: when there is a cancellation between the two terms. And a clear notion of naturalness: when ${\cal V}_\alpha(\phi)$ is comparable to  $W(\phi)$. The same goes for $U_\alpha$, and for their  derivatives, of course. Hence the natural value for $U''_\alpha$ is 
\begin{equation}\label{crit}
U''_\alpha \sim W''|_{\phi_\alpha} \,.
\end{equation}
The right panel of Fig.~\ref{fig4} confirms this picture. If one tunes the brane potentials ${\cal V}_\alpha$ so that $U_\alpha''$ is smaller than the criterion \eqref{crit} then the scalar mode is artificially light. Clearly, because the UV and IR contributions carry warp factors, realizing a light scalar by tuning requires a great deal of tuning on the branes. However without any tuning the UV contribution is generically subdominant ($A_0\equiv 0$) to the others. From now on, we will simply ignore in the discussion the UV brane contribution.

On the other hand, the IR brane contribution can be large without tuning ${\cal V}_{IR}$, provided $\beta$ is small at the IR brane location, so that in that case the dilaton is incarnated by the location of the IR brane. This is the realization that has been discussed in Refs.~\cite{Coradeschi:2013gda,Bellazzini:2013fga}~\footnote{In this dilaton incarnate, its mass scales like  
$m_{\rm dil}^2 \sim \beta^2_{IR} \Lambda_{IR}^2$ because the IR brane potential is implicitly adjusted to have the special property that the potential vanishes at the minimum.}. In the CFT picture, it is interpreted as the condensation threshold of another CFT operator $\cal O'$ in a sudden limit where ${\rm Dim}({\cal O}')$ is arbitrarily large. For this reason, we shall refer to this realization as a {\em hard dilaton}. In the extra dimension literature this dilaton is often called the radion.

The final case to realize a light dilaton is like in Ref.~\cite{Megias:2014iwa} which is when $\Pi_2$ is large. The conditions for which this can happen were analyzed in~\cite{Megias:2014iwa} and require:  \textbf{i)} That $\beta$ is large (of order 1) at the deep IR, and; \textbf{ii)} That it displays a sharp increase just before getting large. Let us call the scale at which this happens $\Lambda_{IR}$. In the presently discussed models, $\Lambda_{IR}\sim m_{KK}$. Under conditions i) and ii), the integral  scales like $\Pi_2^{-1}\sim  \Lambda_{IR}^2 \beta_{IR}^{(-)}$ with $\beta_{IR}^{(-)}$ the (small) value of $\beta$ at $\Lambda_{IR}$ and $m_{\rm dil}^2 \sim \beta_{IR}^{(-)} \Lambda_{IR}^2$. In that case, the dilaton does not correspond to the IR brane location but rather the location of the jump in $\beta(\phi)$, which can be identified as the condensation threshold of the CFT operator dual to $\phi$. For these reasons, we shall refer to this realizations as a {\em soft dilaton}.

Interestingly enough, the dilaton seems to be incarnated by whichever is the largest condensate -- which makes sense from the CFT point of view. Indeed, if $\Pi_2$ is large the presence of the IR brane does not alter much the picture if it is located on the large $\beta$ region and ${\cal V}_{IR}$ is not tuned. Conversely, if the IR brane is stabilized in the region where $\beta$ is small, then the piece of the geometry corresponding to the ${\cal O}$ condensation threshold is cut away and the $\Pi_2$ term cannot be large. 

\subsubsection{Vectors and tensors}

\underline{\textit{Vectors}}

\noindent In the following we will refer to the spectrum of vector perturbations as KK gauge bosons in the 5D warped model.  The gauge bosons are computed from the 5D action of the gauge field $V_M$
\begin{equation}
S =  - \frac{1}{4}\int d^5x \sqrt{-g}  F_{MN}F^{MN}  \,, \label{eq:Svector}
\end{equation}
where $F_{MN}$ is the 5D field strength of the bulk gauge boson $A_M$. In order to compute the mass of massive KK modes we can safely neglect electroweak breaking (as we are assuming the latter will modify the mass of KK modes by a tiny amount) and will make no distinction between KK modes of photons and gluons, and KK modes of $W$ and $Z$ gauge bosons.

\begin{figure}[htb]
\vskip .5cm 
 \begin{tabular}{cc}
 \includegraphics[width=7.5cm]{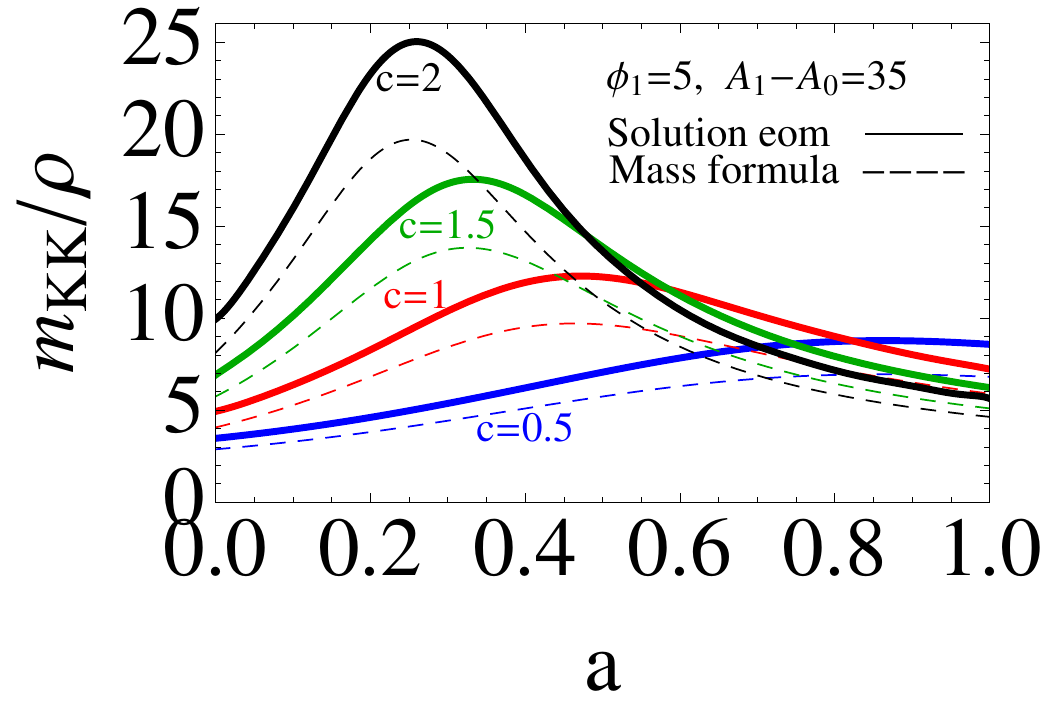}   &
 \includegraphics[width=7.5cm]{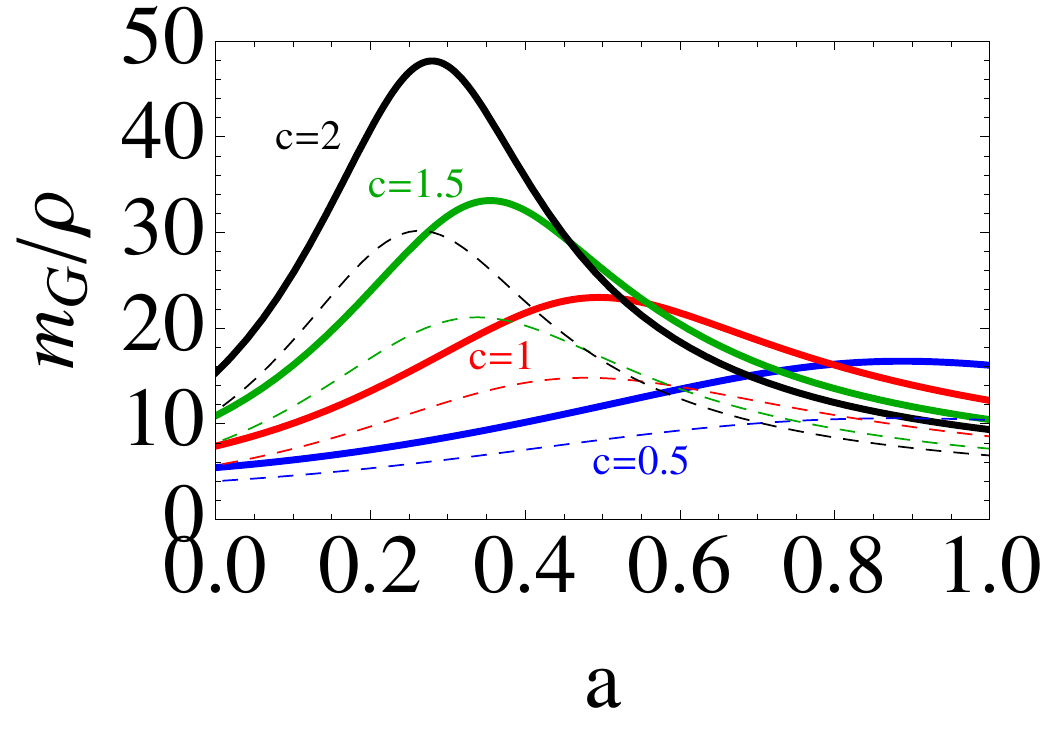}          
\end{tabular}
\caption{\it 
Left panel: KK gauge boson masses as a function of the parameter $a$. Right panel: KK graviton masses as a function of $a$. In these figures we have plotted the results for $c=0.5, 1, 1.5$ and $2$, and have considered $\phi_1 = 5$. We display as wide (solid) lines the results from a numerical computation of the equations of motion, Eqs.~(\ref{gaugeEOM}) and (\ref{eq:eomtensor}), and as narrow (dashed) lines the result from the mass formulae, Eqs.~(\ref{eq:mfvector}) and (\ref{eq:mftensor}).}
\label{fig:vectortensor}
\end{figure}

One can make the KK-mode expansion ansatz 
\begin{equation}
A_\mu(x,y)=  f_A(y)\cdot A_\mu(x)/\sqrt{y_1}
\label{descomp}
\end{equation}
 by which the profiles $f_A$, normalized as $\int_0^{y_1}f^2_A(y)dy=y_1$, and with Neumann boundary conditions $\dot{f}_A(y_\alpha)=0$, satisfy the equations of motion (see e.g.~Ref.~\cite{Ponton:2012bi} for details on the derivation)
\begin{equation}
m_A^2f_A+\left( e^{-2A}\dot{f}_A \right)^{\dt{}}=0  \,.
\label{gaugeEOM}
\end{equation}
We show in Fig.~\ref{fig:vectortensor} (left panel) the result of the mass of the lighest (non-zero) KK gauge boson, $m_{KK}$, as a function of the parameter~$a$. Contrary to the dilaton mass, the KK gauge mode tends to a finite value in the limit $a \to 0$, in particular $m_{KK}/\rho \simeq 2.45 \times 2^{c}$. This is in agreement with the expectations from AdS~\footnote{Note that in the limit $a\to 0$ the superpotential, Eq.~(\ref{nuestro}), tends to the constant value $W(\phi) = 6 k\times 2^c$. There is a factor~$2^c$ affecting the cosmological constant, and hence it affects also the KK modes in this limit.}.

\noindent\underline{\textit{Tensors}}

\noindent Finally the spectrum of tensor perturbations, which we refer in the following as KK modes of the graviton, can be computed by considering the transverse traceless fluctuations of the metric
\begin{equation}
ds^2 = dy^2 + e^{-2A} \left( \eta_{\mu\nu} + h_{\mu\nu}(x,y) \right)dx^\mu dx^\nu  \,.
\end{equation}
Using the ansatz $h_{\mu\nu}(x,y) = h_{\mu\nu}(x)\cdot \vartheta(y)$ one can obtain the equation of motion for the fluctuations $\vartheta(y)$, which is given by
\begin{equation}
\ddot \vartheta - 4 \dot A \dot \vartheta + e^{2A} m^2 \vartheta=0 \,, \label{eq:eomtensor}
\end{equation}
supplemented with Neumann boundary conditions $\dot \vartheta(y_\alpha)=0$ in the UV/IR branes. See e.g.~Ref.~\cite{Cabrer:2009we} for further details. The mass of the lighest (non-zero) KK graviton mode, $m_G$, as a function of the parameter~$a$ is displayed in Fig.~\ref{fig:vectortensor} (right panel). The KK graviton masses have a similar behavior as the KK gauge masses. However, we find that the former are heavier than the latter, in particular they tend to the value $m_{\textrm{G}}/\rho \simeq 3.83 \times 2^c$ in the limit $a \to 0$. A heavier value for the graviton mass with respect to the KK gauge mass was also found in Ref.~\cite{Cabrer:2009we}.\\

\noindent\underline{\textit{Mass formulae}}

\noindent Let us finally mention that, similarly as for the dilaton, it is possible to obtain mass formulae for the first KK gauge boson and graviton in presence of both UV and IR branes, and with full inclusion of the backreaction. The formulae read
\begin{equation}
\frac{\rho^2}{m_{KK}^2} \simeq \frac{ k^2\,
\int_{0}^{y_1}dy \int_{y}^{y_1}dy^\prime e^{2(A-A_1)} \int_{y^\prime}^{y_1} dy^{\prime\prime} 
}{\int_{0}^{y_1} dy} \,, \label{eq:mfvector}
\end{equation}
and
\begin{equation}
\frac{\rho^2}{m_{\textrm{G}}^2} \simeq \frac{k^2\, 
\int_{0}^{y_1}dy e^{-2A}\int_{y}^{y_1}dy^\prime e^{4(A-A_1)} \int_{y^\prime}^{y_1} dy^{\prime\prime} e^{-2(A-A_1)} 
}{\int_{0}^{y_1} dy e^{-2A}} \,, \label{eq:mftensor}
\end{equation}
for the KK gauge and graviton modes, respectively. The analytical approach that allows to obtain these mass formulae predicts, in addition to these non-zero modes, the existence of massless modes in both cases: $m^2_{KK}=0$ (in the absence of electroweak symmetry breaking) and $m^2_{\textrm{G}}=0$ (for the graviton zero mode). We show in Fig.~\ref{fig:vectortensor} the behavior of the first non-zero KK gauge and graviton modes, computed with these mass formulae. It is remarkable that the mass formulae reproduce the numerical evaluation of the equations of motion with an accuracy of $20\%$ for these modes which are not particularly light.

\section{Electroweak breaking}
\label{sec:EWPT}

We will now introduce the electroweak sector in the theory. We shall consider a 5D version of the Standard Model propagating in the 5D space described above. There are several reasons to do so, rather than assuming that the SM is localized on the IR brane. The first one has to do with flavor -- the peculiar flavor structure of the SM then boils down to how different flavor of fermions are localized in the extra dimension. 
Another reason has to do with the dilaton properties, especially for the {\em soft} dilaton case, which is perhaps the one that admits more variability. In the soft case, the dilaton wave function is delocalized in the extra dimension. If this has to have a chance to act as a Higgs impostor, then the SM matter fields better be also in the bulk. Finally, as we will see, a Higgs slightly delocalized from the IR brane is highly favored by EWPT.

Let us emphasize that we also include a 5D Higgs field in the bulk as the source of EW breaking. This, however, does not necessarily mean that we have a SM Higgs in the light spectrum. The 5D Higgs parameters could be such that there is no light `zero' mode in this 5D field, similarly to Higgsless models, or simply that the Higgs zero mode is heavy enough. In particular for the dilaton impostor application, we will assume this kind of situation. 

Thus, we define the 5D $SU(2)_L\times U(1)_Y$ gauge bosons as $W^i_M(x,y)$, $B_M(x,y)$, where $i=1,2,3$ and $M=(\mu,5)$, and the 5D SM Higgs as
\be
H(x,y)=\frac{1}{\sqrt 2}e^{i \chi(x,y)} \left(\begin{array}{c}0\\h(y)+\xi(x,y)
\end{array}\right)
\ ,
\label{Higgs}
\ee
where the matrix $\chi(x,y)$ contains the three 5D SM Goldstone fields $\vec\chi(x,y)$. The Higgs background $h(y)$ as well as the metric $A(y)$ will be considered for the moment as arbitrary functions. Thus the 5D action for the gauge fields and the Higgs field $H$ is written as
\be
S_5=\int d^4x dy\sqrt{-g}\left(-\frac{1}{4} \vec W^{2}_{MN}-\frac{1}{4}B_{MN}^2-|D_M H|^2-V(H)\right)
\label{5Daction}
\ee
where $V(H)$ is the 5D Higgs potential.

Electroweak symmetry breaking will be triggered on the IR brane. We choose the Higgs dependent bulk and brane potentials as
\be
V(H)=M^2(\phi)|H|^2,\quad \mathcal V^0(H)=M_0|H|^2, \quad \mathcal V^1(H)=-M_1|H|^2+\gamma|H|^4  \,.
\ee
The background Higgs field is then determined from the Higgs bulk potential $V(h)=\frac{1}{2}M^2(\phi)h^2$ as
\be
\ddot{h}(y)-4\dot{A} \dot{h}(y)-M^2(\phi)h(y)=0,
\quad \dot{h}(y_\alpha)=\left.\frac{\partial \mathcal V^\alpha}{\partial h}\right|_{y_\alpha} \,.
\label{Higgsec}
\ee
As for the Higgs mass bulk term we will take~\cite{Cabrer:2011fb}
\be
M^2(\phi)=\alpha k\left[\alpha k-\frac{2}{3}W(\phi)  \right]
\label{M2}
\ee
where the parameter $\alpha$ is constrained by the hierarchy problem~\footnote{Notice that for the AdS metric this requires $\alpha>2$.}. The choice (\ref{M2}) ensures that one linearly independent solution to Eq.~(\ref{Higgsec}) is given by $\exp(\alpha ky)$. We can then write the general solution to (\ref{Higgsec}) as
\be
h(\phi)=h(\phi_0)e^{\alpha k y(\phi)}\left[1+(M_0/k-\alpha) Q(\phi)\right],\quad Q(\phi)=\int_{\phi_0}^\phi \left[e^{4A(\bar\phi)-2\alpha k y(\bar\phi)} /W'(\bar\phi)\right]d\bar\phi  \,. \label{eq:hphi}
\ee
To avoid the fine-tuning required in $M_0/k-\alpha$ to keep the exponential solution we must generically require the function $Q(\phi)\lesssim \mathcal O(1)$. As the integrand in $Q(\phi)$ is a monotonically increasing function of $\phi$, a sufficient condition is that 
\be
\alpha=2A(y_1)/y_1  \,.
\label{valordealpha}
\ee
The value of the parameter $\alpha$ can be easily read off from the right panel of Fig.~\ref{fig3} where we plot contour lines of $\alpha/2$ in the plane $(a,c)$. We see that in all cases $\alpha>2$.
There is a simple holographic interpretation as the dimension of the Higgs condensate operator $\mathcal O_H$ depends on the coordinate value~$y$, or equivalently on the value of the field $\phi$, see Ref.~\cite{Cabrer:2011fb}. 

The formalism of electroweak breaking by the bulk Higgs was widely described in Ref.~\cite{Cabrer:2011fb}. The profiles for the massive zero modes of $A_\mu(x,y)$ ($A=W,Z$) are given by $f_A(y)$ as in Eq.~(\ref{descomp}) and satisfy the equation of motion:
\begin{equation}
m_A^2f_A+\left( e^{-2A}\dot{f}_A \right)^{\dt{}}-M_A^2f_A=0
\label{gaugefields}
\end{equation}
where~\footnote{The 5D gauge coupling $g_5$ is related to the 4D one $g$ by $g_5=g\sqrt{y_1}$.} 
\begin{equation}
M_W(y)=\frac{g_5}{2}h(y)e^{-A(y)},\quad M_Z(y)=\frac{1}{c_W}M_W(y) \,.
\end{equation}
In the 4D theory the physical degrees of freedom are the gauge fields $W_\mu(x)^\pm, Z_\mu(x)$, the Goldstone bosons $G_W^\pm(x), G_Z(x)$, and gauge invariant scalars $K_W^\pm(x), K_Z(x)$  which are normally much heavier than the gauge bosons and can thus be considered as decoupled from the low energy effective theory. In case the lightest mode after electroweak breaking is separated by a gap from the KK spectrum we can write for the zero modes the approximations~\cite{Cabrer:2011fb}
\begin{align}
m_A^2&\simeq\frac{1}{y_1}\int_0^{y_1}M_A^2(y) dy \,, \nonumber\\
f'_A(y)&\simeq m_A^2 y_1e^{2A(y)}\left( \Omega(y)-\frac{y}{y_1}\right) \,,
\label{aproximacionfprime}
\end{align}
where
\be
\Omega(y)=\frac{\omega(y)}{\omega(y_1)},\quad \omega(y)=\int_{0}^{y}
h^2(\bar y)e^{-2A(\bar y)} d\bar y \,.
\ee

To compare the model predictions with electroweak precision tests (EWPT) a convenient parametrization is using the $(S,T,U)$ variables in Ref.~\cite{Peskin:1991sw}. They can be given the general expressions~\cite{Cabrer:2011fb}
\begin{align}
\alpha T&=s^2_W \frac{m_Z^2}{\rho^2}k^2 y(\phi_1)\int_{\phi_0}^{\phi_1}\left\{
\left[1-\Omega(\phi)\right]^2e^{2A(\phi)-2A(\phi_1)}/W'(\phi)\right\} d\phi
\nonumber\\
\alpha S&=8c^2_Ws^2_W \frac{m_Z^2}{\rho^2}k^2 y(\phi_1)\int_{\phi_0}^{\phi_1}\left\{
\left(1-\frac{y(\phi)}{y(\phi_1)}\right)\left[1-\Omega(\phi)\right]e^{2A(\phi)-2A(\phi_1)}/W'(\phi)\right\} d\phi\nonumber\\
\alpha U&\simeq 0
\end{align}
where $\rho=ke^{-A(y_1)}$.

We have compared our benchmark class of models in Eq.~(\ref{nuestro}) with EWPT using the fitted values for the $S$ and $T$ parameters as~\cite{PDG}:
\be
T=0.05\pm 0.07,\quad S=0.00\pm 0.08\qquad \textrm{(90\% correlation)}
\ee
and study compatibility with experimental data for values $m_{KK}\leq 10$ TeV and $m_{\rm dil}\leq 1.2$ TeV. 
\begin{figure}[htb]
\vskip .5cm 
 \begin{center}
 \includegraphics[width=7.5cm]{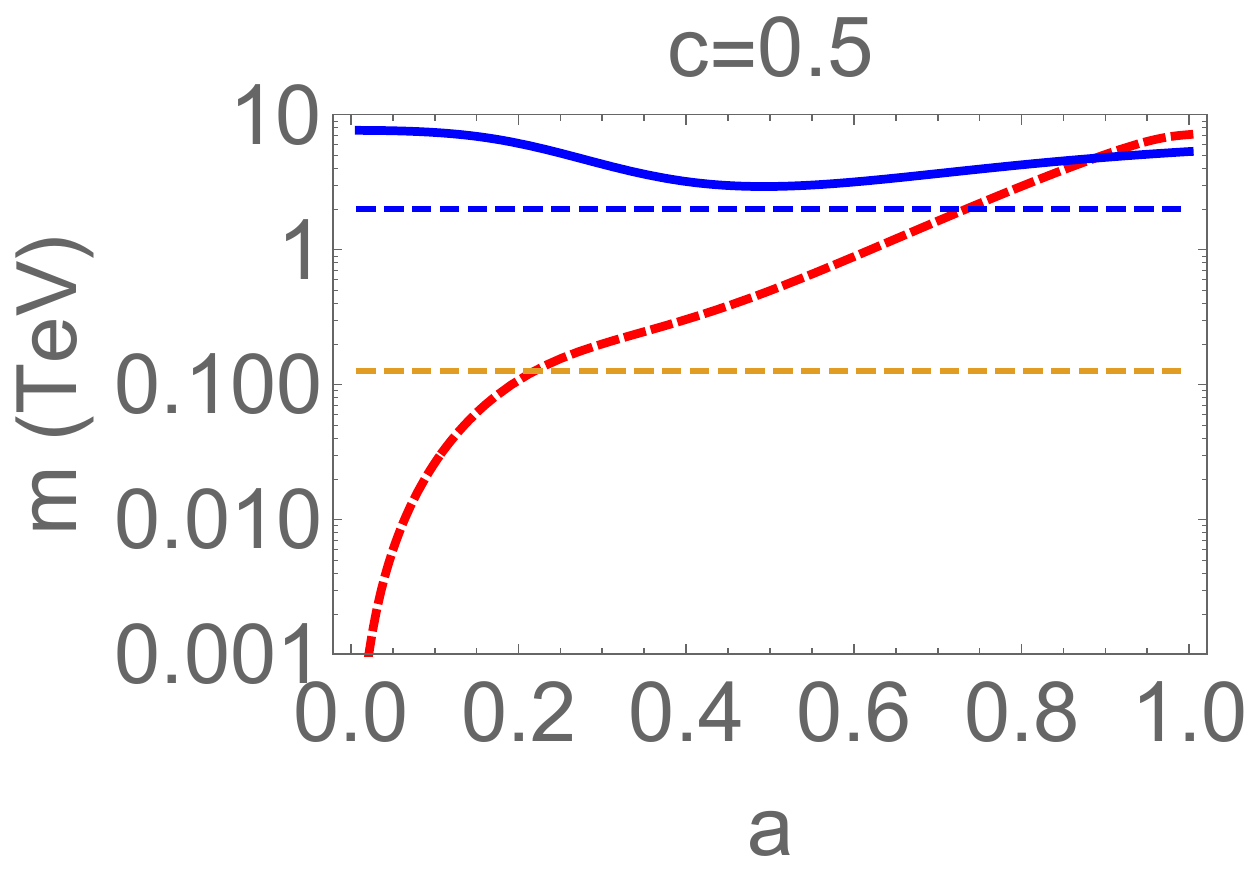}  
 \includegraphics[width=7.5cm]{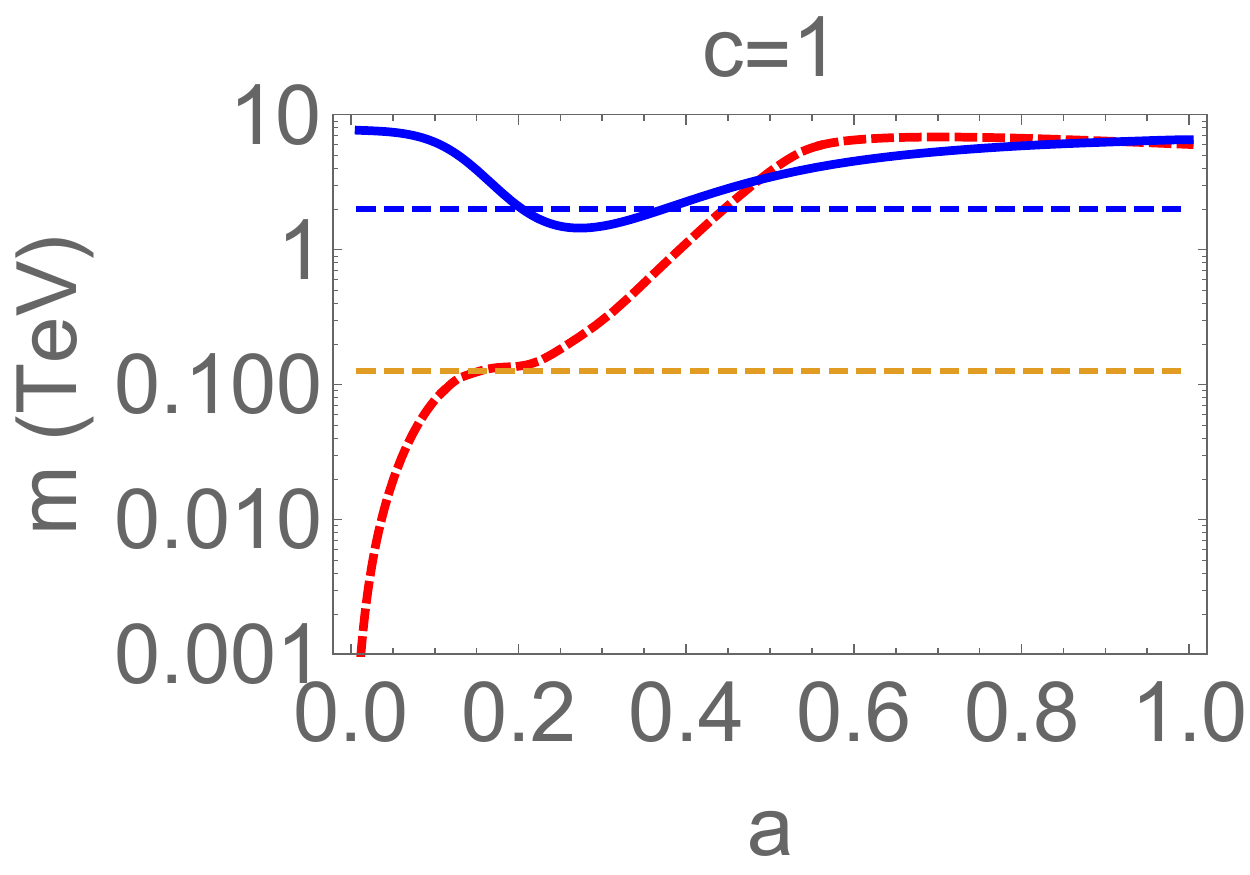}   

 \includegraphics[width=8.3cm,height=5.3cm]{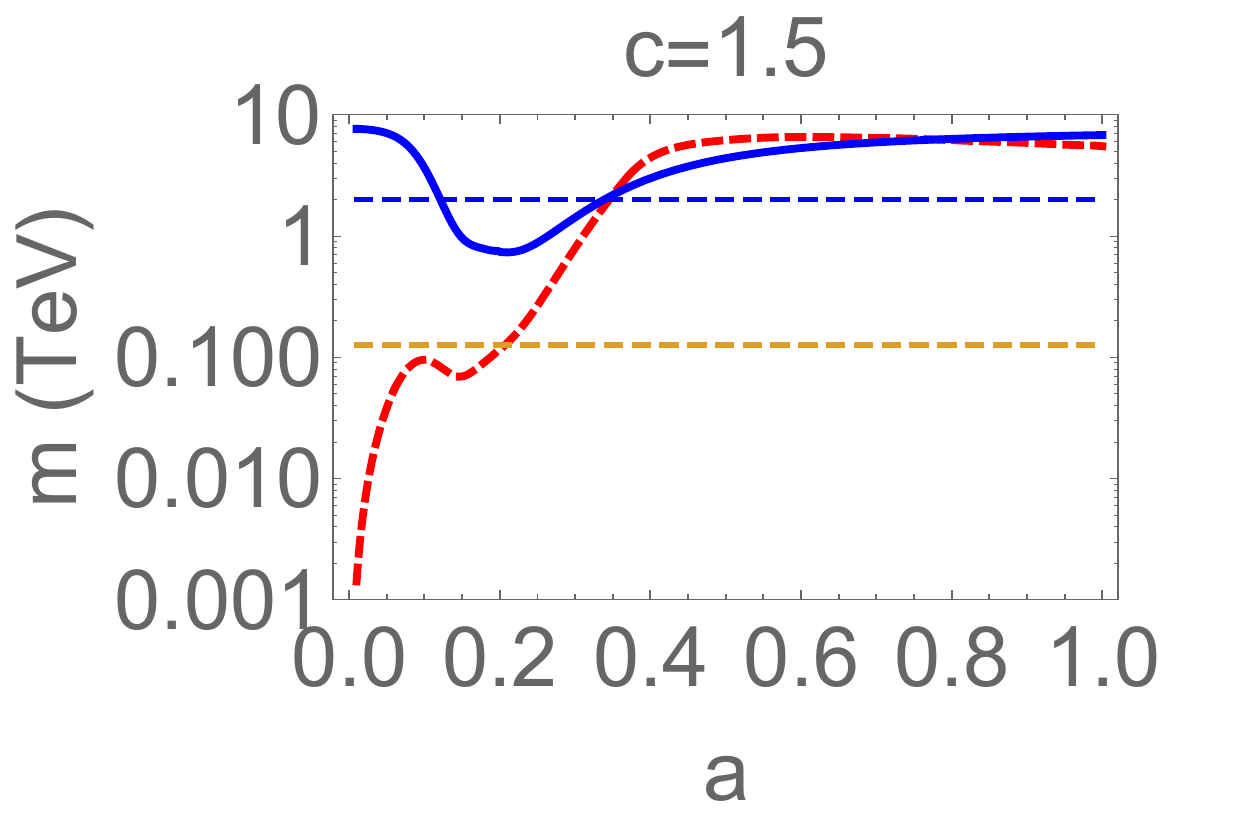} 
 \includegraphics[width=7.5cm]{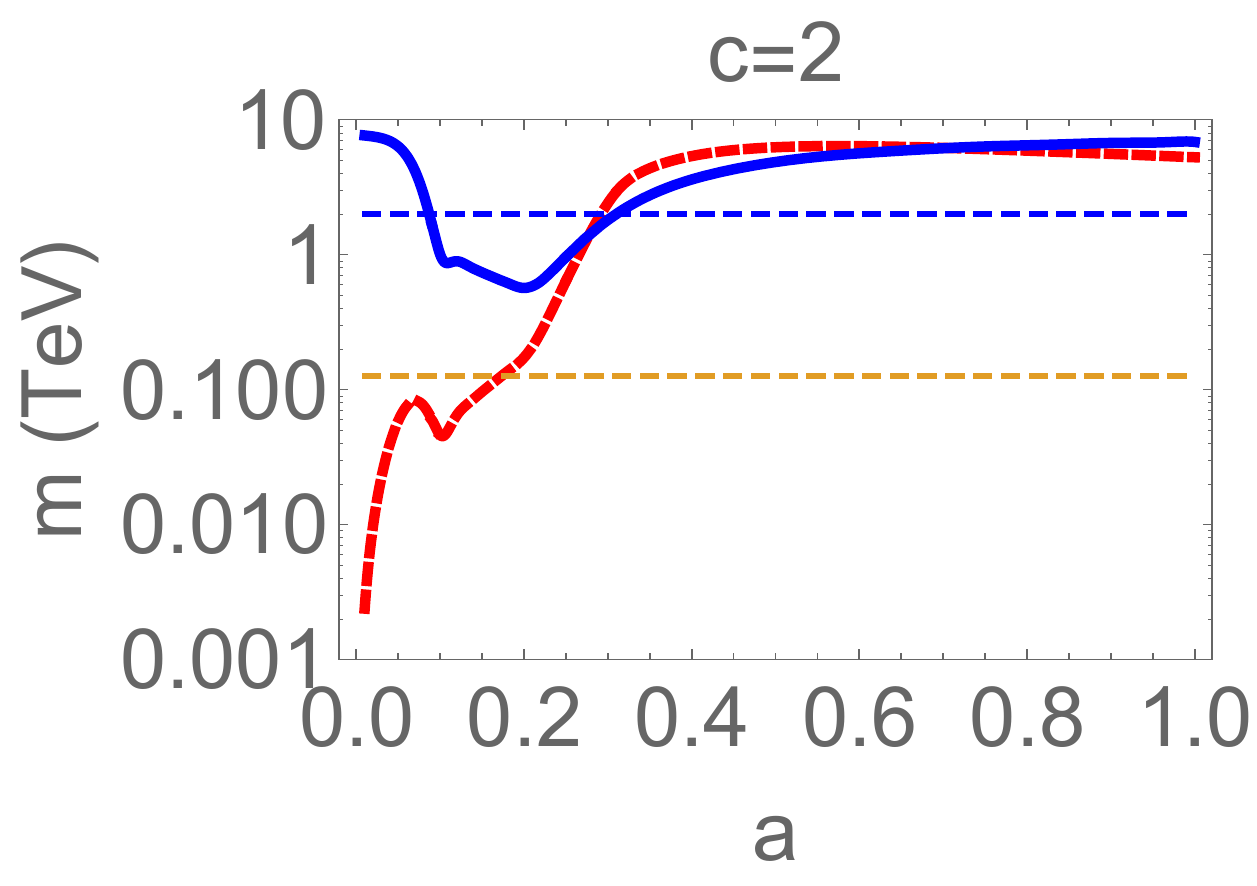}    

\end{center}
\caption{\it Bounds on KK masses (blue solid lines) as functions of $a$ for different values of $c$ from electroweak observables. The corresponding dilaton masses are in dashed (red) lines. To guide the eye we have drawn horizontal dashed lines corresponding to 125 GeV (lower dashed) and 2 TeV (upper dashed).}
\label{fig6}
\end{figure}
The results are summarized in Fig.~\ref{fig6} where we plot the lower bounds on the first KK mode gauge mass (solid blue curve) and the dilaton mass (dashed red curve) as a function of the $a$ parameter for the cases $c=0.5$ (upper left panel), $c=1$ (upper right panel), $c=1.5$ (lower left panel) and $c=2$ (lower right panel). We also show for the sake of reference the horizontal lines corresponding to $m_{KK}=2$ TeV (dashed blue line) and to $m_{\rm dil}=125$ GeV (dashed orange line).  

In the limit of large values of $a$ we recover the Randall-Sundrum (RS)~\cite{Randall:1999ee} case with $\alpha=2$ as can be shown in Fig.~\ref{fig3}. So for large values of $a$ we essentially recover the bounds we would have obtained for the RS case with a bulk Higgs and $\alpha=2$.
In the limit $a\to 0$ we see that an equivalent to the RS case is found where $k\to 2^c k$ which implies that $\alpha\to 2^{c+1}\alpha$ and thus a Higgs more localized toward the IR brane than in RS and consequently bounds on KK masses stronger, as we can  clearly see from the plots in Fig.~\ref{fig6}. As the theories tend to RS (with different values of $\alpha$) in both limits of $a$ large and small, implying large values of $m_{KK}$, $\mathcal O(10)$ TeV, there is a minimum in between for a range of values of the parameter $a$ where the bound on $m_{KK}$ is in the (few) TeV region. This interval and the minimum of $m_{KK}$ scales like $\sim 1/c$ and thus moves towards smaller values of $a$ for larger values of $c$. On the other hand the dilaton mass scales with $a$ so that a lighter dilaton is favored for smaller values of the parameter $a$. In this way moving towards larger values of $c$ (e.g.~$c=2$) favors finding a region in $a$ such that both $m_{KK}=\mathcal O$(TeV) and $m_{\rm dil}\lesssim\mathcal O$(100) GeV, as we can see in the lower right panel of Fig.~\ref{fig6}.

\section{Dilaton couplings in the minimal model}
\label{sec:coupling}

There are many studies in the literature on the coupling of the radion and KK tower of scalar modes to SM matter fields, see e.g.~\cite{Csaki:2000zn,Csaki:2007ns}. It was originally assumed that the matter fields were localized in the IR brane. While this may be considered a priori a reasonable assumption, the situation nowadays is not so clear. There are some recent works that explore the possibility of localizing the matter fields in the bulk, or even in the UV brane, see e.g. Refs.~\cite{Yang:2012dd,Xie:2014loa}. In this work, and motivated mainly by EWPT, we are assuming that the matter and Higgs fields are in the bulk. We study in this section the coupling of the radion with gauge bosons and fermions.

The scalar perturbation of the background flow solution writes as in Eq.~(\ref{dilatonexcitations}). To study the coupling of the radion to matter, one has to firstly find the canonically normalized radion. For that we have to compute the normalization of its kinetic term in the action. If the fluctuation decomposes as $F(x,y) =F(y) \mathcal{R}(x)$, where $F(x,y)$ is dimensionless, the kinetic part of the Lagrangian density in Eq.~(\ref{action}) then reads as
\begin{equation}
M^3\int_{0}^{y_1} dy\sqrt{-g} \left( R -\frac{1}{2}(\partial_M \phi)^2 \right)  \supset M^3 (\partial_\mu \mathcal{R})^2 X_F
\end{equation}
with
\begin{equation}
X_F \equiv \int_{0}^{y_1} dy \left[ 6 + \frac{36^2}{2\beta^2 W^2} \left( \frac{\dot F}{F} - 2 {\dot A} \right)^2 \right] e^{-2A(y)} F^2(y) \,. \label{eq:kineticFgauge}
\end{equation}
The second term inside the bracket in this equation follows from the term proportional to $(\partial_M \phi)^2$ in the action~\footnote{This term turns out to be a small correction to the first term in Eq.~(\ref{eq:kineticFgauge}) for small values of the dilaton mass. In fact, the wave function of a massless mode $m^2\simeq 0$ is $F_{0}(y) \simeq e^{2 A(y)}$, and in this case this term is vanishing.}. If we denote the canonically normalized radion field as $r(x)$ with kinetic term $\frac{1}{2} (\partial_\mu r)^2$, then one finds
\begin{equation}
r(x) = M_{Pl} \left(\frac{X_F}{ \int_{0}^{y_1} dy\, e^{-2A(y)}}\right)^{1/2}  \mathcal{R}(x) \,, \label{eq:r}
\end{equation}
where $M_{Pl}=2.4\times 10^{18}$ GeV and we have made use of Eq.~(\ref{relacion}). 

\subsection{Coupling to gauge bosons}

We will compare the dilaton coupling with massless and massive gauge bosons, and fermions, with the SM Higgs coupling given by
\begin{equation}
\mathcal L_{SM}\supset\frac{\alpha_{EM}}{8\pi}(b_1+b_{1/2})\frac{h_{SM}}{v}F_{\mu\nu}F^{\mu\nu} -\frac{h_{SM}}{v}\left(2 m_W^2 W_\mu W^\mu+m_Z^2 Z_\mu Z^\mu+m_f \bar f f\right)\,,
\end{equation}
where $b_1=-7$, $b_{1/2}=(4/3)^2$, $F_{\mu\nu}$ is the photon field strength and $v=246$ GeV. Now in order to compute the coupling of the dilaton with matter we will expand the 5D action (\ref{5Daction}) to linear order in the perturbations in Eq.~(\ref{dilatonexcitations}). We will make extensive use of the formalism developed in Ref.~\cite{Cabrer:2011fb} which we have summarized in Appendix~\ref{sec:AppendixA}.  In case the Higgs is in the bulk (as we are assuming in this paper) we can compute the coupling of the dilaton to the photon, massive gauge fields ($W_\mu$ and $Z_\mu$) and fermions $f$ normalized with respect to the SM Higgs couplings as
\be
\mathcal L_{rad}=\frac{r(x)}{v}\left\{ \frac{\alpha_{EM}}{8\pi}(b_1+b_{1/2})\,c_\gamma\,F_{\mu\nu}F^{\mu\nu}-2 c_W\,m_W^2 W_\mu W^\mu-c_Z\,m_Z^2 Z_\mu Z^\mu-c_f m_f \bar f f \right\}  \label{eq:LV}
\ee
where of course the case $c_\gamma=c_W=c_Z=c_f=1$ corresponds to the SM Higgs coupling.

The linearized 5D action for gauge bosons reads as
\begin{align}
\delta_F S_5&=\int d^4xdy F(x,y)\left\{-\frac{1}{2}F_{\mu\nu}F^{\mu\nu}-2e^{-2A} \left[ (Z_{\mu 5})^2+2|W_{\mu 5}|^2  \right]\right.\nonumber\\
&\left.
+\frac{2}{y_1}e^{-2A}\left[M_Z^2(y) \eta_Z^2(y) K_Z^2(x) +2 M_W^2(y) |\eta_W(y)|^2 |K_W(x)|^2  \right] 
\right\}
\label{deltaF}
\end{align}
where $Z_{MN}$ and $W_{MN}$ refer to the 5D field strengths of $Z_M$ and $W_M$, respectively, and the last line refers to mass terms for pseudoscalars which the dilaton also couples to.  Notice that the third term $|D_\mu H|^2$ in the action (\ref{5Daction}) does not contribute to the dilaton coupling to matter as there is an accidental cancellation~\footnote{In case the Higgs is localized in the IR brane the localized Higgs kinetic term does provide the usually considered coupling of the radion to the gauge boson masses. As we are considering here the Higgs propagating in the bulk of the fifth dimension even if such coupling can be generated by radiative corrections it should be suppressed by loop factors and we are not considering it here.}. A straightforward calculation gives the 4D Lagrangian
\begin{align}
\mathcal L_{rad-VV}&=-\mathcal R(x)\frac{1}{2}F_{\mu\nu}^2(x) \frac{\int_0^{y_1}F(y)}{y_1}-2y_1\int_0^{y_1}dy\,F(y)e^{2A(y)}\left(\Omega-\frac{y}{y_1}\right)^2\nonumber\\
&\cdot \mathcal R(x)\left[ m_Z^2(\partial_\mu G_Z(x)-m_Z Z_\mu(x))^2+2 m_W^2|\partial_\mu G_W(x)-m_W W_\mu(x)|^2  \right]
\label{Lrad}
\end{align}
where use has been made of the approximation Eq.~(\ref{aproximacionfprime}). 

\begin{figure}[htb]
\vskip .5cm 
 \includegraphics[width=7.8cm,height=6cm]{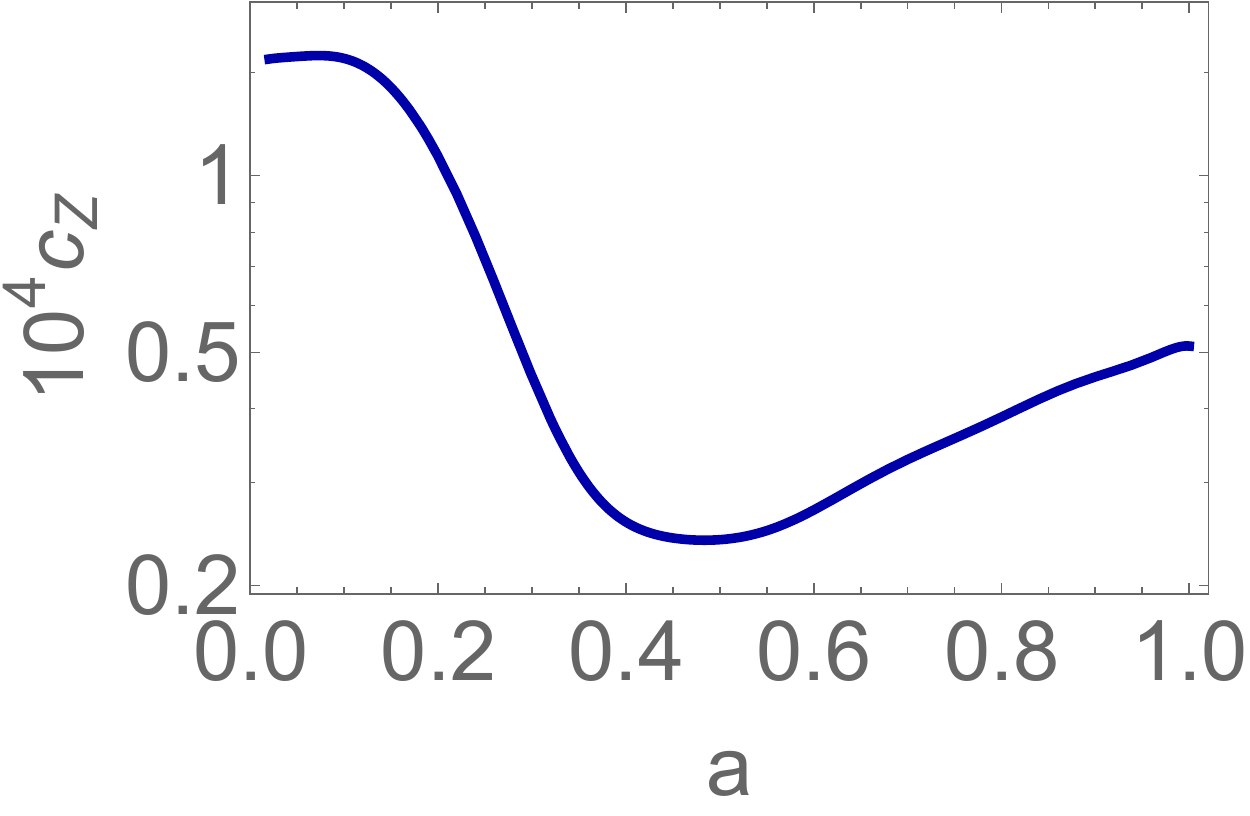}  
 \includegraphics[width=7.8cm,height=6cm]{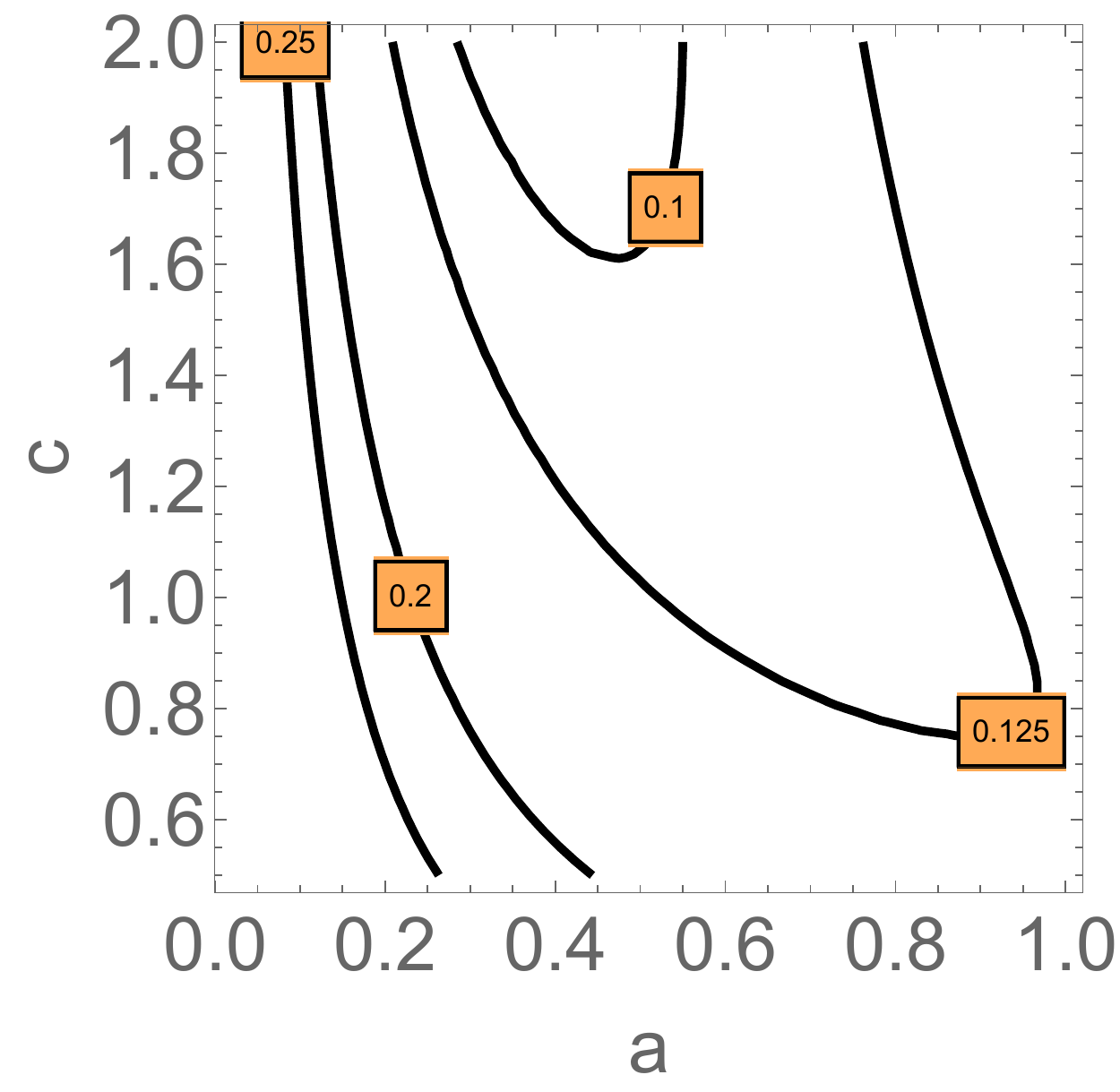}    
\caption{\it Left panel: Plot of $c_Z$ as a function of $a$ for $c=1$. Right panel: Contour lines for constant values of $c_\gamma$.}
\label{fig7}
\end{figure}

Equation~(\ref{Lrad}) provides a small coupling of the radion field to electroweak gauge bosons, as the first term is volume suppressed~\footnote{A suppression $\sim 10-35$ depending on the values of the parameters $a$ and $c$.} and the second term has (with respect to the case where the Higgs is localized on the IR brane) an extra suppression of $m_V^2/\rho^2$. We can then write $c_\gamma$ and $c_V$ $(V=W,Z)$ as
\begin{align}
c_\gamma&=-\frac{4\pi}{\alpha_{EM}(b_1+b_{1/2})}\frac{v}{M_{Pl}}\frac{\int dy\,F}{y_1}\left(\frac{\int dy\,e^{-2A}}{X_F}\right)^{1/2} \,, \nonumber \\
c_V&=2\,\frac{v\, m_V^2}{M_{Pl}}\left(\frac{\int dy\,e^{-2A}}{X_F}\right)^{1/2}y_1\int dy\,F e^{2A}\left(\Omega-\frac{y}{y_1}\right)^2  \,,
\end{align}
or using the massless radion approximation $F=e^{2A}$
\begin{align}
c_\gamma&=-\frac{4\pi}{\alpha_{EM}\sqrt{6}(b_1+b_{1/2})}\frac{v}{e^{-A_1}M_{Pl}}\left(k^2\int dy\, e^{-2A}\int dy\, e^{2A-2A_1}  \right)^{1/2}\frac{1}{ky_1} \,, \nonumber\\
c_V&=\frac{2\,v\, m_V^2}{\sqrt{6} e^{-A_1}M_{Pl} \rho^2}\left(\frac{\int dy\, e^{-2A}}{\int dy\, e^{2A-2A_1}}  \right)^{1/2} ky_1 \int_0^{ky_1}dy\,e^{4A-4A_1}\left(\Omega-\frac{ky}{ky_1}  \right)^2 d(ky)  \,.
\end{align}
We plot in the left panel of Fig.~\ref{fig7} the value of $c_Z$ as a function of $a$ for $c=1$ (left panel) after imposing the constraints from EWPT of Sec.~\ref{sec:EWPT}. 
In the right panel of Fig.~\ref{fig7} we plot contour lines of constant $c_\gamma$ in the plane $(a,c)$. We discuss in Sec.~\ref{sec:dilext} what are the implications of these results. 

\subsection{Coupling to fermions}

\begin{figure}[htb]
\vskip .5cm 
\begin{center}
 \includegraphics[width=9.5cm]{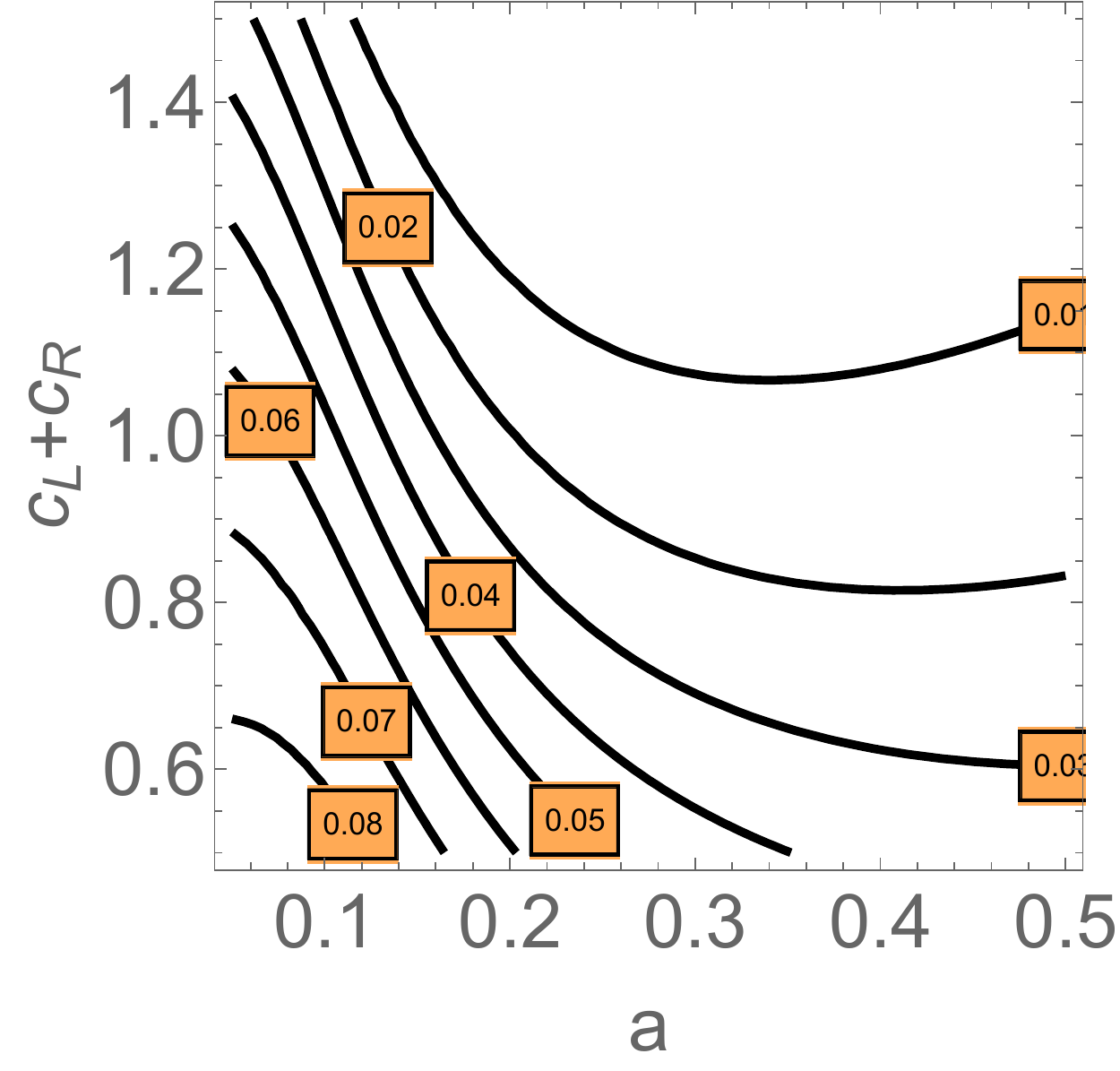}  
\end{center}
\caption{\it Contour lines of $c_f$ for the case $c=1$ in the plane $(a,c_L+c_R)$.}
\label{fig8}
\end{figure}

The couplings of the radion with the SM fermions is model dependent and should depend on the 5D Dirac masses which determine the localization of the different fermions. The latter were fixed in Ref.~\cite{Cabrer:2011qb}  to $M_{f_{L,R}}(y)=\mp c_{L,R}W(\phi)$ where the upper (lower) sign corresponds to the left (right) component. With this convention light (heavy) fermions are localized near the UV (IR) brane and have $c_{L,R}>1/2$ ($c_{L,R}<1/2$). Using appropriate boundary conditions one can have for every 5D Dirac fermion one massless zero mode corresponding to a given chirality, 
\be
\psi^{(0)}_{L,R}(y,x)=\frac{e^{(2-c_{L,R})A(y)}}{\left(\int dy\, e^{A(1-2 c_{L,R})} \right)^{1/2}} \psi_{L,R}(x)
\ee
and null wave function for the opposite chirality ($\psi^{(0)}_{R,L}(y)\equiv 0$). The effective 4D fermion Lagrangian for the radion interaction with fermion zero modes is then written as:
\be
\mathcal L_{rad-f f}=-3m_f\bar\psi_L(x)\psi_R(x)\mathcal R(x) 
\frac{\int dy\, e^{\alpha y-(c_L+c_R)A}F(y)}{\int dy\ e^{\alpha y-(c_L+c_R)A}}
\ee
and correspondingly the normalized coupling $c_f$ is given by
\be
c_f=3 \frac{v}{M_{Pl}}\left(\frac{\int dy\,e^{-2A}}{X_F}  \right)^{1/2}\frac{\int dy\,e^{\alpha y-(c_L+c_R)A}F(y)}{\int dy\,e^{\alpha y-(c_L+c_R)A}}
\ee
We show in  Fig.~\ref{fig8} how this coupling depends on the model parameters. 

Of course when we build a complete theory of flavor by different localization in the bulk (Dirac masses) of different SM (5D) fermions we have to worry not only at oblique (universal) observables but also at non-oblique ones, as e.g.~$R_b$, as well as at FCNC and CP violating operators. This was done for the benchmark model with $c=1$ in Ref.~\cite{Cabrer:2011qb} where it was proven that the lower bound on $m_{KK}$ increases from $\mathcal O$(TeV) to around $\gtrsim 3$~TeV at least. As working out a complete theory of flavor is largely beyond the scope of this paper we will disregard possible bounds on non-universal and flavor observables just keeping in mind that in a more realistic theory the latter issue should be confronted with experimental data.

\subsection{Radion-Higgs mixing}
\label{sec:mixing}
In theories where the Higgs propagates in the bulk, there is an extra effect by which the radion (a bulk field) mixes with the Higgs. In fact we have seen in the previous section that there is no radion coupling with the Higgs kinetic term $-|D_\mu H|^2$ as there was an accidental cancellation. However the radion does couple to the terms in the action $-|D_5 H|^2-V(H)$, which 
\begin{figure}[htb]
\vskip .5cm 
\begin{center}
 \includegraphics[width=11.5cm]{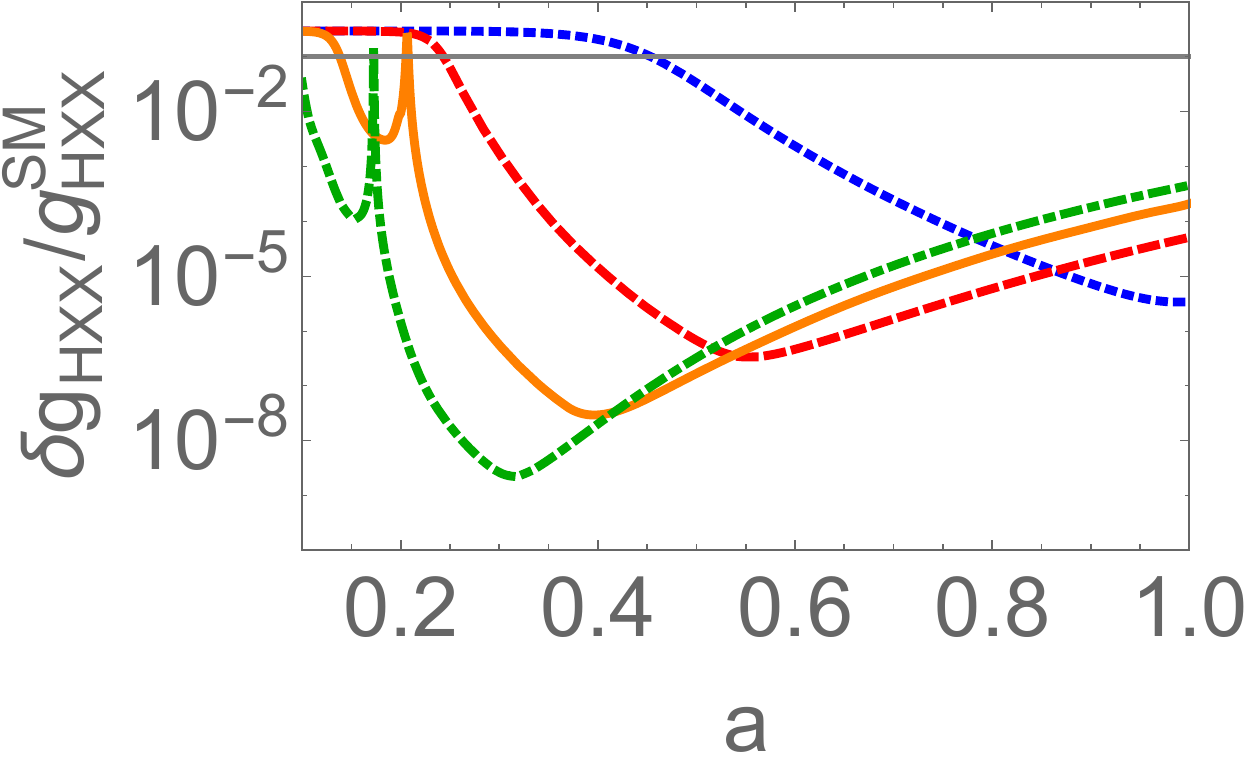}  
\end{center}
\caption{\it The value of $\delta g_{HXX}/ g_{HXX}^{SM}$ generated from the radion-Higgs mixing as a function of the parameter $a$ for different values of $c=0.5$ (blue dotted), 1 (red dashed), 1.5 (orange solid)  and 2 (green dot-dashed).}
\label{fig9}
\end{figure}
yields a coupling of the radion with the Higgs field and in particular a mixing between the radion  and the Higgs as $\mathcal L_{rH}= \mu^2\, r(x) \mathcal H(x)+\dots$ where $r(x)$ and $\mathcal H(x)$ are the canonically normalized radion and Higgs fields.
A straightforward calculation yields the result
\be
|\mu^2|=\rho^2 \left(\frac{2\int e^{-2A}}{3\int e^{2(A-A_1)}}  \right)^{1/2}\frac{v}{e^{-A_1}M_P}
\frac{\int e^{-2A+2\alpha k y}(3\alpha^2+M^2/k^2)}{\int e^{-2A+2\alpha k y}}
\label{mix}
\ee
where we are considering the light dilaton approximation $F(x,y)=e^{2 A(y)}\mathcal R(x)$.

As a consequence of the mixing in Eq.~(\ref{mix}) the radion and Higgs states are rotated by an angle $\beta$. The two mass eigenstates are projected to the Higgs $\mathcal H(x)$ with coefficients $\sin\beta$ and $\cos\beta$ respectively. For small mixing (as we will assume) the Higgs-like eigenstate will couple with SM fields $XX$ through $\mathcal H(x)$ with coupling $g_{HXX}=\cos\beta g_{HXX}^{SM}$ so that the mixing with the radion generates a shift in the coupling  $\delta g_{HXX}=(1-\cos\beta) g_{HXX}^{SM}$. Present data from LHC provide limits on the Higgs couplings which can be taken as $\delta g_{HXX}/ g_{HXX}^{SM}\lesssim 0.1$. The values of $\delta g_{HXX}/ g_{HXX}^{SM}$ stemming from the radion-Higgs mixing are shown in Fig.~\ref{fig9} as a function of the parameter $a$ for different values of $c=0.5$ (blue dotted), 1 (red dashed), 1.5 (orange solid)  and 2 (green dot-dashed). The grey solid line indicates the upper bound $\delta g_{HXX}/ g_{HXX}^{SM}= 0.1$. It translates into lower bounds on the parameter $a$ which depend on the value of $c$. In particular $a\gtrsim 0.4$ for $c=0.5$, $a\gtrsim 0.25$ for $c=1$ while $a$ is essentially unbounded for the cases with $c=1.5$ and $c=2$. The sharp peaks which appear in the latter cases for isolated values of $a$ arise from a resonance effect when $m_{dil}=m_H=125$ GeV. 

There is also a radion-Higgs mixing from the brane localized potentials $\mathcal V_\alpha(\phi)$ but its value is negligibly small and will not be considered in this paper. From here on we will assume we are in a region of the parameter space where the mixing is negligible

\subsection{Standard Model plus a light dilaton}
\label{sec:dilext}

The above results show that the numerical values of $c_V$ and $c_f$ are $\ll 1$ 
in the `minimal' model where the 5D SM fields couple minimally to the 5D metric as in \eqref{5Daction}. This has two implications: first of all, possibilities like a Higgs impostor cannot be covered by the present model.  
Second of all and for the same reason, in models with a light dilaton {\em in addition} to the full SM (Higgs included), the dilaton is quite weakly coupled. In particular, the Higgs phenomenology (Higgs contribution to the unitarization of the $V_LV_L$ elastic and inelastic scattering, the Higgs strengths, etc.) will be affected by an $\mathcal O(10^{-4})$ effect.
The tiny deviations with respect to the SM predictions would be unobservable at the LHC.

\section{The diphoton excess}
\label{sec:diphoton}

As the radion couples to photons and gluons by dimension five operators as in
\be
\mathcal L_{rad-VV}=-\frac{1}{2 y_1}\mathcal R(x)(F_{\mu\nu}^2+G_{\mu\nu}^2)\int_0^{y_1} F(y)+\dots
\label{lagrangianoradion}
\ee
where $F_{\mu\nu}$ and $G_{\mu\nu}$ are the field strengths for photons and gluons, respectively, an immediate question arises. Can the radion be responsible for the recent $\gamma\gamma$ excess found at ATLAS and CMS~\cite{excess}?

\begin{figure}[htb]
\vskip .5cm 
\begin{center}
 \includegraphics[width=10cm]{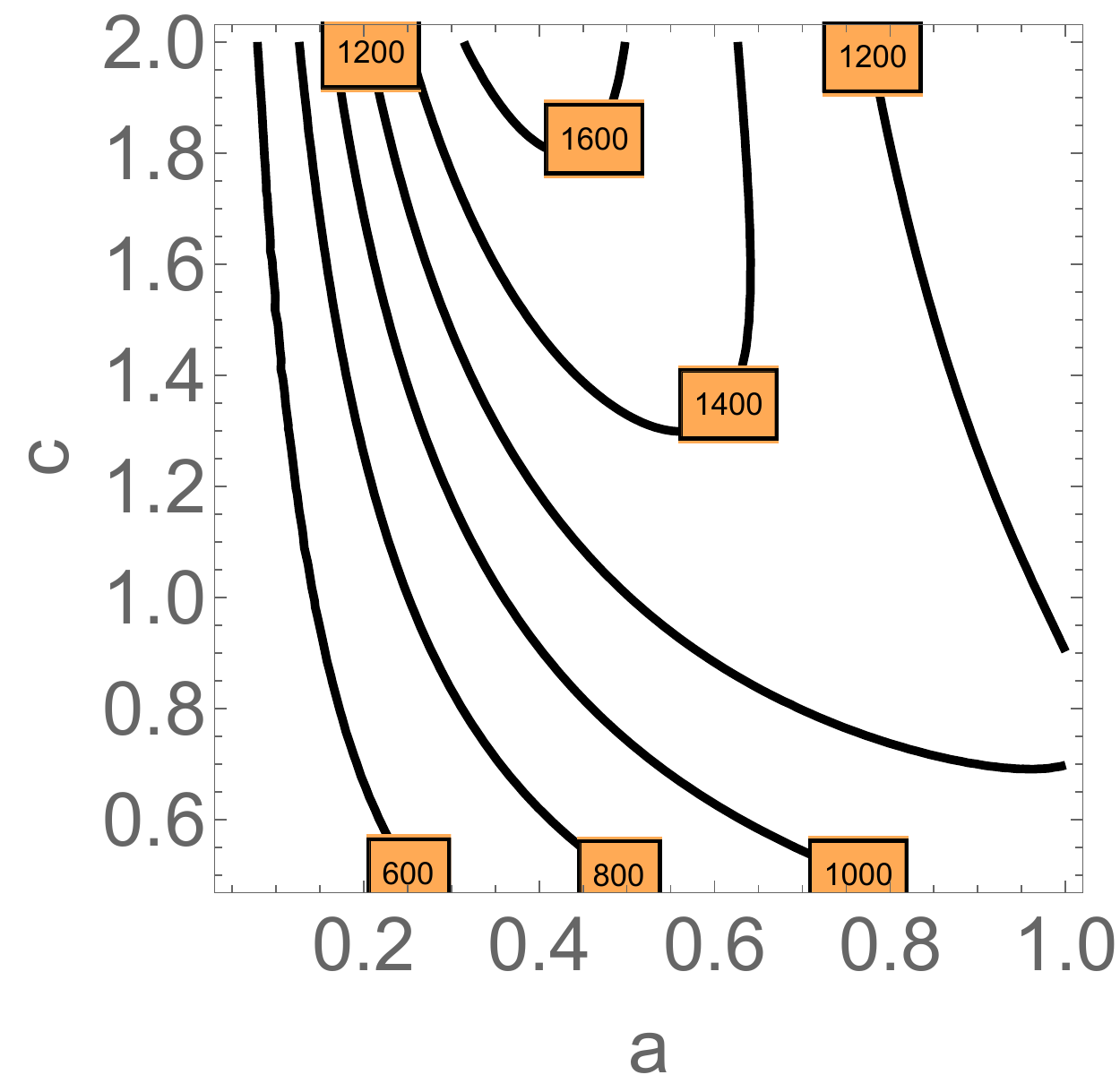}  
\end{center}
\caption{\it Contour lines of constant $\Lambda_\gamma=\Lambda_G$ in TeV.}
\label{fig10}
\end{figure}

To answer this question we will write the Lagrangian (\ref{lagrangianoradion}) as an effective Lagrangian
\be
\mathcal L_{eff}=\frac{1}{\Lambda_\gamma}r(x) F_{\mu\nu}^2(x)+\frac{1}{\Lambda_G}r(x) G_{\mu\nu}^2(x)
\ee
where $r(x)$ is the canonically normalized radion defined in Eq.~(\ref{eq:r}) and the scales $\Lambda_{\gamma,G}$ are given by
\be
\Lambda_\gamma=\Lambda_G=\sqrt{6}e^{-A_1}M_P\left( \int e^{-2A} dy\int e^{2(A-A_1)}dy\right)^{-1/2}
\ee
Contour lines of $\Lambda_\gamma=\Lambda_G$ in TeV, in the plane $(a,c)$, which parametrize our models, are presented in Fig.~\ref{fig10}.

The analysis of the values of $\Lambda_{\gamma,G}$ required to explain the $\gamma\gamma$ excess can be found in Ref.~\cite{Franceschini:2015kwy}~\footnote{A general analysis also including $\Lambda_W$ can be found in Ref.~\cite{Cao:2015pto}.}. Assuming only the channels $r\to\gamma\gamma$ and $r\to gg$ the condition for the production of a resonance of mass $M=750$ GeV and total width $\Gamma_t/M \simeq 0.06$ is~\footnote{The correlation between $\Lambda_\gamma$ and $\Lambda_G$ has also been studied in Ref.~\cite{Chakrabortty:2015hff}.}
\begin{align}
&\frac{M^2}{\Lambda_\gamma^2}+8\frac{M^2}{\Lambda_G^2}\simeq 0.75\nonumber\\
& \frac{M}{\Lambda_\gamma} \frac{M}{\Lambda_G}\simeq 1.35\times 10^{-3}
\end{align}
which yields in particular the solution 
\be
\Lambda_\gamma\simeq 171\textrm{ TeV, } \Lambda_G\simeq 2.44 \textrm{ TeV.} 
\label{valores}
\ee
A quick glance at the plot in Fig.~\ref{fig10} shows that in the minimal theory we have presented in previous sections the radion is too weakly coupled to $\gamma\gamma$ and $gg$ to be able to explain the alleged excess in $\gamma\gamma$ found by ATLAS and CMS.

It is clear that if the radion has to describe the diphoton excess the model has to be modified in some way. The model as defined in \eqref{action} and \eqref{5Daction} is far from generic in the sense that many other operators can be included  with the same field content and symmetries in the 5D Lagrangian. Most importantly, the way how the Golberger-Wise scalar $\phi$ enters the EW sector is clearly over-restrictive. Since we have an important and nontrivial scalar profile $\phi(y)$ that drives the bulk geometry, it is far from clear why should the direct couplings of the SM fields to $\phi$ vanish. 
Equation  \eqref{5Daction} implicitly assumed that the 5D SM fields couple universally to the 5D metric only, so that the couplings to $\phi$ (and thus the dilaton) arise only through the metric. However, this is more a matter of convenience and simplicity than a property that is protected by some symmetry. Thus, we find it natural in this section to include more general couplings. The simplest higher dimensional operator that do not affect the background and can have an impact on the coupling of the radion to photons and gluons is then
\be
\Delta S_5=\int d^5x\sqrt{-g}\left[\mathcal O_B(\Phi) B_{MN}B^{MN}+\mathcal O_G(\Phi) G_{MN}G^{MN}   \right] 
\label{DeltaS}
\ee
where the operator $\mathcal O_{X}(\Phi)$ ($X=B,G$) is given by
\be
\mathcal O_X(\Phi)=\frac{1}{k^2}\left( \partial_M\Phi\partial_M\Phi-W'^2(\Phi)\right) Z_X(\Phi)
\ee
and the function $Z_X$ is a smooth function of $\Phi$ which we will take as an exponential $Z_X(\Phi)=e^{d_X\Phi}$ with $d_X$ a real coefficient. We are thus introducing a real parameter in  the theory in order to fit the excess. We will not try to justify the presence of the higher dimensional operator in Eq.~(\ref{DeltaS}) which should require some UV completion of the theory. Instead we will work out the region in the parameter space where the excess can be explained for a radion with mass $M=750$ GeV. We will see this can be accomplished without any tuning of the theory, and with all parameters of $\mathcal O(1)$.

The additional radion couplings generated by (\ref{DeltaS}) can be obtained by expanding to linear order the prefactor as
\begin{align}
&(\delta_F+\delta_\phi)
\left\{ \left(\partial_M \Phi \partial^M \Phi-W'^2(\Phi)  \right)Z(\Phi)\right\}
=-2F(y)W'^2 Z(\phi) \mathcal R(x)\nonumber\\
&+
4\mathcal R(x) \left[3\ddot F(y)-(W(\phi)+6W''(\phi))\dot F(y)+ (2W(\phi)W''(\phi)-W'^2(\phi))F(y) \right]Z(\phi)\nonumber\\
&=4\mathcal R(x)\left( \frac{1}{2}W'^2-3e^{2A}m^2 \right)F(y)Z(\phi)
\label{deltaFphi}
\end{align}
where $m$ is the radion mass and in the last equality we have made use of the radion equation of motion. We can then write the correction to the 4D Lagrangian as
\be
\Delta \mathcal L_{rVV}=\frac{1}{\Lambda_\gamma}r(x) F_{\mu\nu}^2(x)+\frac{1}{\Lambda_Z}r(x) Z_{\mu\nu}^2(x)+\frac{1}{\Lambda_G}r(x) G_{\mu\nu}^2(x)
\label{esta}
\ee
where
\be
\Lambda_\gamma=\Lambda_B/\cos^2\theta_W,\quad \Lambda_Z=\Lambda_B/\sin^2\theta_W
\ee
and
\be
\Lambda_X=-\left(\frac{3 \int e^{2(A-A_1)dy}}{8\int e^{-2A}dy}  \right)^{1/2}\frac{y_1 e^{-A_1}M_P}
{{\displaystyle \int }\left( \frac{1}{2}\frac{W'^2}{k^2}-3e^{2(A-A_1)} \frac{M^2}{\rho^2}  \right)Z_X e^{2(A-A_1)}dy}
\ee

\begin{figure}[h]
\vskip .5cm 
 \begin{center}
 \includegraphics[width=7.cm,height=5.3cm]{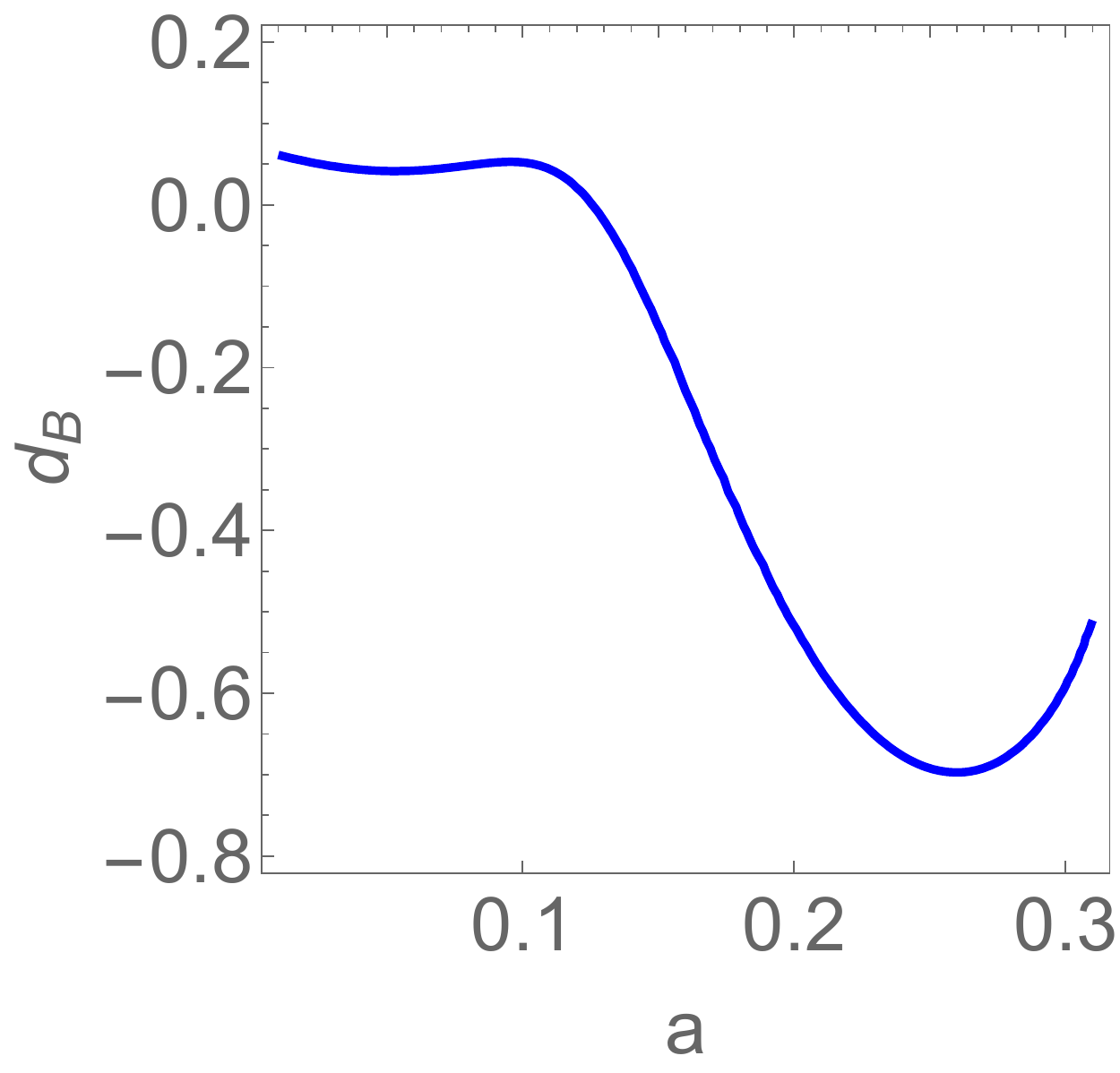}  
 \includegraphics[width=7.cm,height=5.3cm]{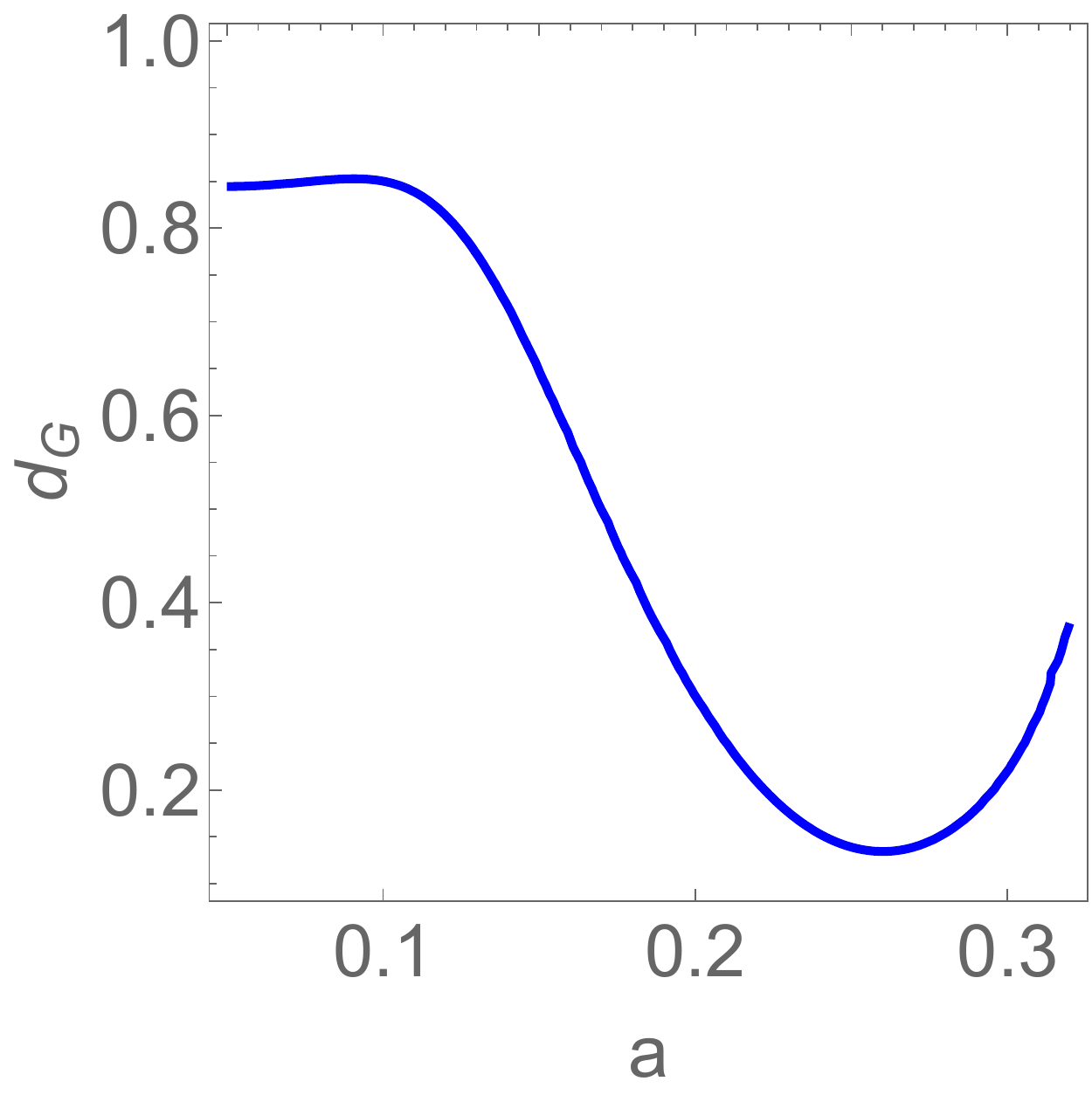}   
 \includegraphics[width=7cm,height=5.3cm]{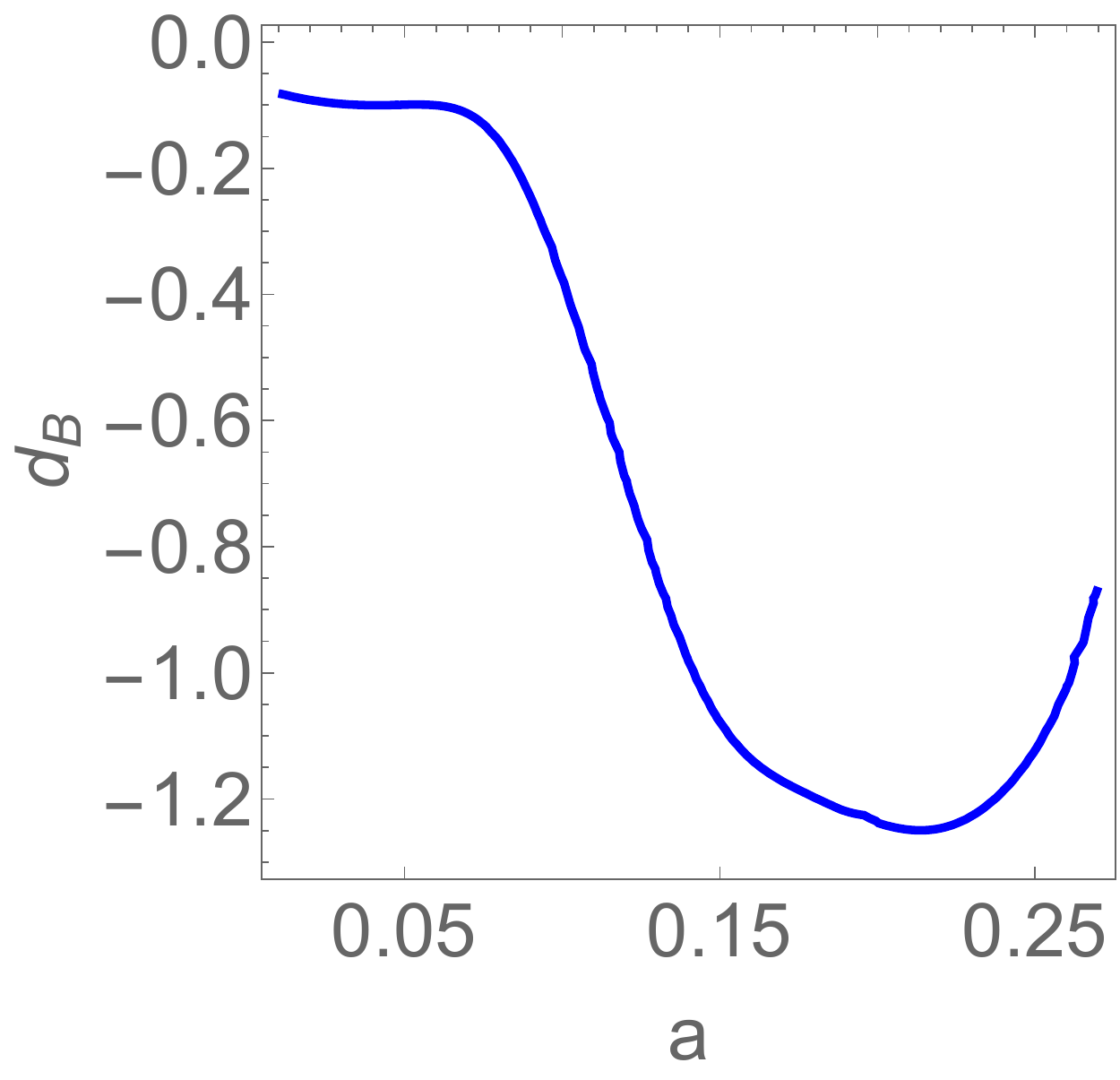} 
 \includegraphics[width=7.cm,height=5.3cm]{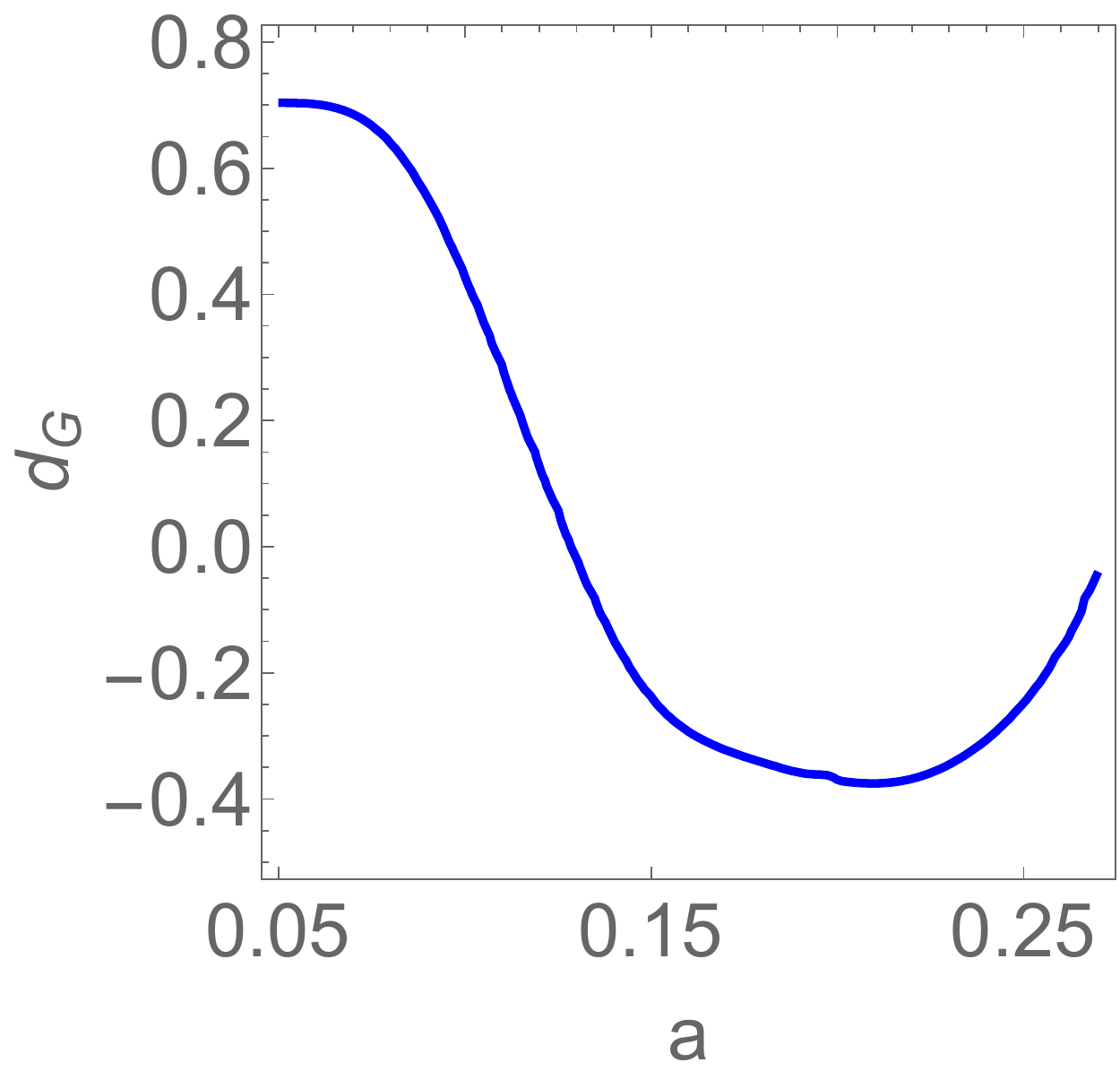}    
\end{center}
\caption{\it Contour plots of $\Lambda_\gamma=171$ TeV (left panels) and $\Lambda_G=2.44$ TeV (right panels), for the cases $c=1$ (upper panels), and $c=1.5$ (lower panels).}
\label{fig11}
\end{figure}

Obviously from the structure of Eq.~(\ref{esta}) we can write the relations~\cite{Franceschini:2015kwy}
\be
\Gamma(r\to ZZ)=\tan^4\theta_W\Gamma(r\to\gamma\gamma),\quad 
\Gamma(r\to Z\gamma)=2\tan^2\theta_W\Gamma(r\to\gamma\gamma)\ ,
\ee
the related decays involving $Z$ bosons are suppressed and  the bounds from the resonant $ZZ$ productions are satisfied. Moreover the excess can be described, as we said above, for values of the effective parameters $\Lambda_\gamma$ and $\Lambda_G$ given in Eq.~(\ref{valores}). In Fig.~\ref{fig11}
we plot contour lines of $\Lambda_\gamma=171$ TeV in the plane $(a,d_B)$ (left panels) and $\Lambda_G=2.44$ TeV in the plane $(a,d_G)$ (right panels), for the cases $c=1$ (upper panels), and $c=1.5$ (lower panels).  We can see that that in all cases we can fix the excess for values of the parameters $|d_{B,G}|$ of $\mathcal O(1)$. We have fixed for all points the dilaton mass to $M=750$ GeV. We can see that the values of the dilaton mass in the plots of Fig.~\ref{fig6}, where we were using $U''(\phi_1)\simeq W''(\phi_1)$, are usually below 750 GeV. However this can be easily fixed by increasing the value of $U''(\phi_1)$, which can be done without any tuning of the parameters, as it was shown in the left panel of Fig.~\ref{fig4}.  Another possibility, of course,  would be to keep $U''(\phi_1)\simeq W''(\phi_1)$ and instead increase the KK scale above $\sim 2$ TeV, but we shall not consider it as it leads to less interesting phenomenology.

\section{Conclusions}
\label{sec:conclusions}

In this work we have revisited the `soft-wall' models with one warped extra dimension, and we have assessed their model-building capabilities towards realizing a light dilaton, understood as the Goldstone boson associated to spontaneously broken conformal invariance. We have identified the key ingredients present in the 5D models that allow to realize dilatons that are\textit{ i)} naturally light and/or \textit{ii)} phenomenologically viable for various distinct applications. We have focused on two main applications: 

1) {\em A dilaton extension of the SM}, where the dilaton is the first new state in the spectrum in addition to all the  Standard Model particles (including the SM Higgs). 

2) {\em A diphoton resonance}, that is a $\sim 750$ GeV mass dilaton state as responsible for the recently found diphoton excess.

Option 1) is, of course, the simplest and least contrived to realize with 5D models. Interestingly enough, experimental data are compatible with such a dilaton being considerably light (below $\sim 100$ GeV), which basically comes about because the dilaton--SM couplings tend to be slightly suppressed.  The strongest lower bound on the dilaton mass in this case comes from the naturalness requirement. The dilaton mass $m^2$ can be naturally lower than the scale of SBCI $\Lambda^2_{KK}$ by a factor given by the value of the $\beta_\Lambda$ function at $\Lambda_{KK}$. In RG-flows that start at a finite UV scale $\Lambda_{UV}$ and realize SBCI naturally, $\beta_\Lambda$ is parametrically suppressed like $\beta_\Lambda \sim 1/\log(\Lambda_{UV}/\Lambda_{KK})$ \cite{CPR}. Taking $\Lambda_{UV}\sim M_{Pl}$ and $\Lambda_{KK}\sim 2$ TeV, one obtains that $m_{\rm dil}$  can be naturally in the  $\sim20$--$50$ GeV range, but not much smaller (without additional mechanisms). Of course the possibility that the dilaton mass is around 125 GeV, and the electroweak breaking Higgs much heavier, opens up the possibility of a dilaton as a Higgss-impostor. However the analysis of the couplings of the 125 GeV dilaton to gauge bosons and fermions highly disfavors this possibility.

Option 2)  is considerably more challenging to realize.  Our results show that it is possible to build well defined five-dimensional models that realize a moderately heavy (750 GeV) dilaton which could explain the diphoton excess recently found at the LHC. However, these models are less minimal and slightly tuned. 
Specifically, the couplings to gauge bosons have to be arranged to accommodate the diphoton excess found at LHC. 
These values do not follow from a symmetry breaking pattern anymore because in these models the gauge bosons break explicitly (by being in the bulk) conformal invariance strongly enough to have no prediction on what their couplings to the dilaton are. Note that this is still compatible with the dilaton being a pseudo-Goldstone boson of SBCI~\footnote{This is clear in the models that we considered because the dilaton is indeed lighter than the KK scale at tree level. It seems that quantum effects in the bulk should not spoil this property, but a proper analysis of this question requires a separate study.}. Note also that we are lead to this kind of model in order to be able to pass the EWPTs. The price to pay for phenomenological viability, then, is that this class of dilaton models does not make a definite prediction on the dilaton couplings to `matter'. Accordingly, the model parameters have to be tuned. 
In our view, though, it seems a bit premature to rule out that there might exist other realizations of a dilaton whose couplings to fermions and gauge bosons follow from the SBCI and at the same time are compatible with experimental data. Again, a 750 GeV dilaton diphoton is not dictated by symmetries, but it can be realized consistently.

Admittedly, there are many questions that have to be left for the future, such as concerning the consistency, naturalness of our effective 5D treatment or the possibility to UV-complete it. But at the very least our construction can be taken as a concrete and working model of a dilatons that (so far) passes all experimental tests and be used to accommodate the recent diphoton excess found at LHC.

\section*{\sc Note added}

While completing this manuscript some references \cite{Bellazzini:2015nxw,Gupta:2015zzs,Cox:2015ckc,Ahmed:2015uqt} where posted on arXiv where the possibility of a 750 GeV radion to explain the LHC diphoton excess was studied. In particular the results from Ref.~\cite{Bellazzini:2015nxw} disfavor a dilaton from the spontaneous breaking of conformal invariance as the source of the diphoton excess, in agreement with the general results from our section~\ref{sec:coupling}. The presence of the operator (\ref{DeltaS}) permits stronger coupling of the dilaton to gauge bosons and seems to spoil the general conclusions of Refs.~\cite{Bellazzini:2015nxw,Gupta:2015zzs}. Their  analyses are more restrictive because  implicitly they do not allow significant additional explicit breaking of conformal invariance in the `matter' sector. These result in additional dilaton couplings that are certainly model-dependent but very important in order for the model to be viable. As emphasized in footnote~16, they do not necessarily interfere with the scalar being a pseudo-Goldstone boson. In Ref.~\cite{Cox:2015ckc} they succeed to explain the diphoton excess but giving up to explain the UV/IR full hierarchy, while in our work we insist in solving the gauge hierarchy as an initial requirement. In Ref.~\cite{Ahmed:2015uqt} the scenario of an IR brane localized Higgs with the radion-Higgs mixing stemming from the localized action $\mathcal R^{(4)}|H|^2$ and a gauged custodial symmetry in the bulk is considered. However in our non-custodial models a localized Higgs is disfavored by EWPT and this was the main motivation to consider a bulk propagating Higgs.

\section*{\sc Acknowledgments}

The work of O.P. and M.Q.~is partly supported by the Spanish
Consolider-Ingenio 2010 Programme CPAN (CSD2007-00042), by MINECO
under Grants CICYT-FEDER-FPA2011-25948 and
CICYT-FEDER-FPA2014-55613-P, by the Severo Ochoa Excellence Program of
MINECO under the grant SO-2012-0234 and by Secretaria d'Universitats i
Recerca del Departament d'Economia i Coneixement de la Generalitat de
Catalunya under Grant 2014 SGR 1450. The work of M.Q. is also partly supported by  CNPq PVE fellowship project 405559/2013-5. E.M. would like to thank the
Institut de F\'{\i}sica d'Altes Energies (IFAE), Barcelona, Spain, the
Instituto de F\'{\i}sica Te\'orica CSIC/UAM, Madrid, Spain, and the Departamento de F\'{\i}sica At\'omica, Molecular y Nuclear of the Universidad de Granada, Spain, for their hospitality during the completion of the final stages of this
work. The work of E.M. is supported by the European Union under a
Marie Curie Intra-European Fellowship (FP7-PEOPLE-2013-IEF) with
project number PIEF-GA-2013-623006.

\section*{Appendix}
\appendix

\section{Standard Model gauge fluctuations}
\label{sec:AppendixA}

We are presenting in this Appendix some further technical details in the computation of Section~\ref{sec:coupling}. In particular the 4D physical degrees of freedom from the 5D gauge fields $[W_M^\pm(x,y), Z_M(x,y)]$ and $[\chi^\pm_W(x,y), \chi_Z(x,y)]$ are the 4D massive gauge fields $[W^\pm_\mu(x),Z_\mu(x)]$, the Goldstone bosons $[G^\pm_W(x),G_Z(x)]$ and the pseudoscalars $[K^\pm_W(x),K_Z(x)]$. 

The decomposition of gauge fields is as in Eq.~(\ref{descomp}) while that of Goldstone bosons and pseudoscalars is given by
\begin{align}
 \sqrt{y_1}A_5(x,y)&=\frac{1}{m_A}G_A(x)\cdot \dt{f}_A(y)-\frac{M_A^2(y)}{m^2_{\eta_A}}K_A(x)\cdot \eta_A(y)\nonumber\\
 \sqrt{y_1}\chi_A(x,y)&=\frac{1}{m_A}G_A(x)\cdot f_A(y)-\frac{1}{m_{\eta_A}^2}M_A^{-2}\left( M_A^2 e^{-2A}\eta_A\right)^{\dt{}}\cdot K_A(x)
\end{align}
where $A=W^\pm,Z$ and $``\cdot"$ indicates expansion over KK modes. The profile $\eta_A(y)$ of the pseudoscalar $K_A(x)$ satisfies the equation of motion with Dirichlet boundary conditions
\begin{equation}
m^2_{\eta_A}\eta_A+\left[ M_A^{-2}\left( M_A^2 e^{-2A}\eta_A\right)^{\dt{}} \right]^{\dt{}}-M^2_A\eta_A=0,\quad \eta_A(y_\alpha)=0
\end{equation}
and the normalization equation
 \begin{equation}
\frac{1}{y_1}\int_0^{y_1}M_A^2e^{-2A}\eta_A^2=1  \,.
\end{equation}
Notice that in the limit $M_A\to 0$ there is no massless mode for $\eta_A$ since the zero mode would have the trivial wave function $\eta_A(y)\equiv 0$ consistent with the Dirichlet boundary conditions. In this case only massive KK modes do appear.  

To quadratic order in the fluctuations, the  5D action for the gauge field $Z_M$ Eq.~(\ref{eq:Svector}) can be written as~\cite{Cabrer:2011fb}
\begin{equation}
S_5= \int d^5x \sqrt{-g} \left( - \frac{1}{4} \left(Z_{\mu\nu}\right)^2 -\frac{1}{2} e^{-2A} (Z_{\mu 5})^2 - \frac{1}{2} M_Z^2 (\partial_\mu \chi_Z - Z_\mu)^2 - \frac{1}{2} M_Z^2 e^{-2A} (\chi^\prime_Z - Z_5)^2  \right)  \label{eq:vectorZ}
\end{equation}
whereas for the $W_M^\pm$ gauge field we have 
\begin{equation}
S_5= \int d^5x \sqrt{-g} \left( - \frac{1}{2} \left|W_{\mu\nu}\right|^2 - e^{-2A} \left|W_{\mu 5}\right|^2 - M_W^2 \left|\partial_\mu \chi_W - W_\mu\right|^2 - M_W^2 e^{-2A} \left|\chi^\prime_W - W_5\right|^2  \right)  \label{eq:vectorW}
\end{equation}

\end{document}